\documentclass[10pt,journal,compsoc]{IEEEtran}

 \usepackage{graphicx}
\usepackage{amsmath,amssymb,amsfonts}
\usepackage{gensymb}
\usepackage{xcolor}

\usepackage{algorithmic}
\usepackage{algorithm}
\usepackage{graphicx}
\usepackage{subcaption}
\usepackage{textcomp}
\usepackage{mathtools}
\usepackage{placeins}

\usepackage{optidef}
\usepackage{amsmath}
\usepackage{textcomp}
\usepackage{makecell}
\usepackage{array}
\usepackage{gensymb}
 \usepackage{multirow,booktabs}
\usepackage{xcolor}
\usepackage{fixltx2e}

\begin{document}

\title{Thermal Prediction for Efficient Energy Management of Clouds using Machine Learning}

\author{Shashikant Ilager,  Kotagiri Ramamohanarao, Rajkumar Buyya%

\author{Shashikant Ilager, Member, IEEE, \\
			 Kotagiri Ramamohanarao, Member, IEEE,
			  Rajkumar Buyya, Fellow, IEEE%
\IEEEcompsocitemizethanks{\IEEEcompsocthanksitem Shashikant Ilager, Kotagiri Ramamohanarao, and Rajkumar Buyya are with The Cloud Computing and Distributed Systems (CLOUDS) Laboratory, School of Computing and Information Systems,
The University of Melbourne, Melbourne, VIC 3010, Australia. (e-mail: shashikant.ilager@gmail.com, \{kotagiri, rbuyya\}@unimelb.edu.au).\protect\\
}}%

\thanks{Manuscript received April 19, 2019; }}
\IEEEtitleabstractindextext{%
\begin{abstract}
Thermal management in the hyper-scale cloud data centers is a critical problem. Increased host temperature creates hotspots which significantly increases cooling cost and affects reliability.  Accurate prediction of host temperature is crucial for managing the resources effectively. Temperature estimation is a non-trivial problem due to thermal variations in the data center. Existing solutions for temperature estimation are inefficient due to their computational complexity and lack of accurate prediction. However, data-driven machine learning methods for temperature prediction is a promising approach. In this regard, we collect and study data from a private cloud and show the presence of thermal variations. We investigate several machine learning models to accurately predict the host temperature. Specifically, we propose a gradient boosting machine learning model for temperature prediction. The experiment results show that our model accurately predicts the temperature with the average RMSE value of 0.05 or an average prediction error of 2.38 $\celsius$, which is 6 $\celsius$ less as compared to an existing theoretical model. In addition, we propose a dynamic scheduling algorithm to minimize the peak temperature of hosts. The results show that our algorithm reduces the peak temperature by 6.5 $\celsius$ and consumes 34.5\% less energy as compared to the baseline algorithm. 
\end{abstract}

\begin{IEEEkeywords}
Cloud computing,  Machine learning, Energy efficiency in a data center, Datacenter cooling, Hotspots
\end{IEEEkeywords}}

\maketitle

\IEEEdisplaynontitleabstractindextext
\IEEEpeerreviewmaketitle
\IEEEraisesectionheading{\section{Introduction}\label{sec:introduction}}

    The transition from ownership-based on-premise IT infrastructure to subscription-based Cloud has been tremendous in the past decade due to the vast advantages that cloud computing offers \cite{buyya2009cloud}. This rapid proliferation of cloud has resulted in a massive number of hyper-scale data centers that generate an exorbitant amount of heat and consume a large amount of electrical energy.  According to  \cite{USreport2016}, around 2\% of global electricity is spent on data centers, and almost 50\% of this energy is spent on cooling systems \cite{patel2003smart}.   
 Modern cloud data centers' rack-mounted servers can consume up to 1000 watts of power each and attain peak temperature as high as 100 $\celsius$ \cite{ASHRAE}.  The power consumed by a  host is dissipated as heat to the ambient environment, and the cooling system is equipped to remove this heat and keep the host's temperature below the threshold.  Increased host temperature is a   bottleneck for the normal operation of a data center as it escalates the cooling cost. It also creates  hotspots that severely affect the reliability of the system due to cascading failures caused by silicon component damage. The report from Uptime Institute \cite{uptime} shows that the failure rate of equipment doubles for every  10 $\celsius$ increase above 21 $\celsius$. Hence, thermal management becomes a crucial process inside the data center Resource Management System (RMS).

Therefore, to minimize the risk of peak temperature repercussions, and reduce a significant amount of energy consumption, ideally, we need accurate predictions of thermal dissipation and power consumption of hosts based on workload level. In addition, a scheduler that efficiently schedules the workloads with these predictions using certain scheduling policies. However, accurate prediction of a host temperature in a steady-state data center is a non-trivial problem \cite{ZhangTPDS2018, tang2008energy}. This is extremely challenging due to complex and discrepant thermal behavior associated with computing and cooling systems.  Such variations in a data center are usually enforced by CPU frequency throttling mechanisms guided by Thermal Design Power (TDP),   attributes associated with hosts such as its physical location, distance from the cooling source, and also thermodynamic effects like heat recirculation \cite{ZhangTPDS2018, tang2008energy}. Hence, the estimation of the host temperature in the presence of such discrepancies is vital to efficient thermal management. Sensors are deployed on both the CPU  and rack level to sense the  CPU  and ambient temperature, respectively. These sensors are useful to read the current thermal status. However, predicting future temperature based on the change in workload level is equally necessary for critically important RMS  tasks such as resource provisioning, scheduling, and setting the cooling system parameters. 
Existing approaches to predict the temperature are inaccurate, complex, or computationally expensive. The widely used theoretical analytical models \cite{ZhangTPDS2018, tang2008energy,  sun2017spatio, Zhang2007cputemp, etas} that are built based on mathematical relations between different cyber-physical components lack the scalability and accurate prediction of the actual temperature. In addition, theoretical models fail to consider several variables that contribute towards temperature behavior and they need to be changed for different data centers.   Computational Fluid Dynamics (CFD) models  are also predominantly used \cite{choi2008cfd, CFDsecond}  for accurate predictions, but their high complexity requires a large number of computing cycles. \textcolor{black}{ Building these CFD models and executing them can take hours or days, based on individual data center complexity  \cite{Moore2006weatherman}. The CFD models are useful in initial design and calibration of data center layout and cooling settings, however, it is infeasible  for the realtime tasks (e.g., scheduling in large scale clouds) that are dynamic and require quick online decisions.} 
\textcolor {black}{Moreover, CFD simulation requires both computational  (e.g, the layout of the Data Center, open tiles) and physical parameters, and changes to these parameters need expensive retraining of the models \cite{zapater2016runtime}. However, our approach is fast and cost-effective as it  solely relies on the physical sensor data that are readily available on any rack-mounted servers and implicitly captures  variations.}  Hence, data-driven methods using machine learning techniques is a promising approach to predict the host temperature quickly and accurately.

Machine learning (ML) techniques have become pervasive in modern digital society mainly in computer vision and natural language processing applications. With the advancement in machine learning algorithms and the availability of sophisticated tools, applying these  ML techniques to optimize large scale computing systems is a propitious avenue \cite{Fox2019, Maolearning, gao2014machine, DRLcloudCheng2018RL}.  Recently, Google has reported a list of their efforts in this direction \cite{JeffMLforSystem}, where they optimize several of their large scale computing systems using ML to reduce cost, energy and increase the performance. Data-driven temperature predictions are highly suitable as they are built from actual measurements and they capture the important variations that are induced by different factors in data center environments. Furthermore, recent works have explored ML techniques to predict the data center host temperature \cite{ZhangTPDS2018, Luo2018}. However,  these works are applied to HPC data centers or similar infrastructure that relies on both application and physical level features to train the models. In addition,  they are application-specific temperature estimations. Nevertheless, the presence of the virtualization layer in Infrastructure clouds prohibits this application-specific approach due to an isolated execution environment provided to users. Moreover, getting access to the application features is impractical in clouds because of privacy and security agreements between users and cloud providers. Consequently, we present a host temperature prediction model that completely relies on features that can be directly accessed from physical hosts and independent of the application counters.

In this regard,  we collect and study data from our University's private research cloud. We propose a data-driven approach to build temperature prediction models based on this collected data. We use this data to build the ML-based models that can be used to predict the temperature of hosts during runtime. Accordingly, we investigated several ML algorithms including variants of regression models,  a neural network model namely Multilayer Perceptron (MLP), and ensemble learning models. Based on the experimental results, the ensemble-based learning, gradient boosting method, specifically,  XGBoost \cite{chen2016xgboost} is chosen for temperature prediction.  The proposed prediction model has high accuracy with an average prediction error of 2.5 $\celsius$ and Root Mean Square Error (RMSE) of 0.05. Furthermore, guided by these prediction models, we propose a  dynamic scheduling algorithm to minimize the peak temperature of hosts in a data center.   The scheduling algorithm is evaluated based on real-world workload traces and it is capable of circumventing potential hotspots and significantly reduces the total energy consumption of a data center. The results have demonstrated the feasibility of our proposed prediction models and scheduling algorithm in data center RMS. 

    In summary, the key contributions of our work are:
    \begin{itemize}
    \item We collect physical-host level measurements from a real-world data center and show the thermal and energy consumption variations between hosts under similar resource consumption and cooling settings.
    \item We build machine learning-based temperature prediction models using fine-grained measurements from the collected data.
    \item We show the accuracy and the feasibility of proposed prediction models with extensive empirical evaluation.
    \item We propose a dynamic workload scheduling algorithm guided by the prediction methods to reduce the peak temperature of the data center that minimizes the total energy consumption under rigid thermal constraints.
    \end{itemize}
    The remainder of the paper is organized as follows:  The motivations for this work and thermal implications in the cloud are explained in Section 2. Section 3  proposes a thermal prediction framework and explores different ML algorithms. Section 4 describes the gradient boosting based prediction model. The feasibility of the prediction model is evaluated against a theoretical model in Section 5. Section 6 presents a dynamic scheduling algorithm. The analysis of scheduling algorithm results is done in Section 7 and the feature set analysis is described in Section 8.  The relevant literature for this work is discussed in Section 9.  Finally, Section 10 concludes the paper and also points out future research directions.

\section{\textcolor{black}{Motivation: Intricacies in Cloud Data Centers' Thermal Management}}\label{sec:motivation}

    \begin{figure}[t]
        \captionsetup{justification=centering}
        \centering
         \begin{subfigure}[t]{0.49\columnwidth}
            \includegraphics[width=\linewidth]{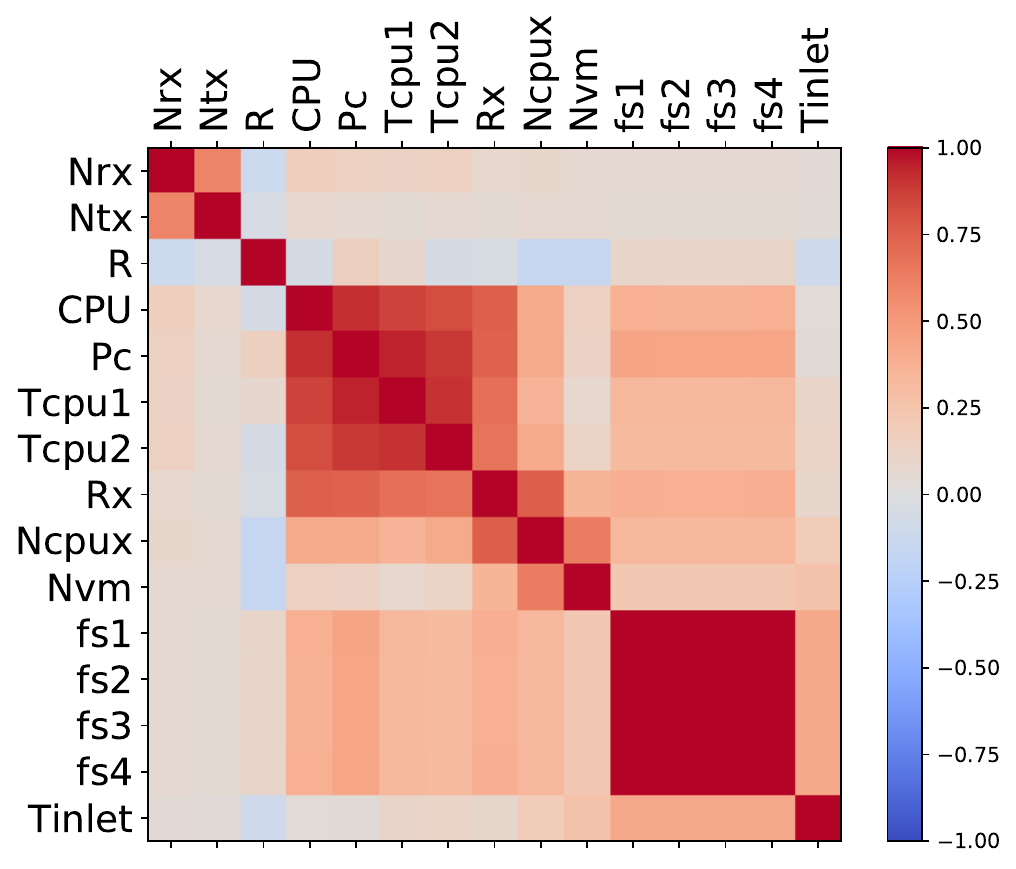}
            \caption{Correlation between all features}
            \label{allcor}
        \end{subfigure} 
        \begin{subfigure}[t]{0.49\columnwidth}
           \includegraphics[width=\linewidth]{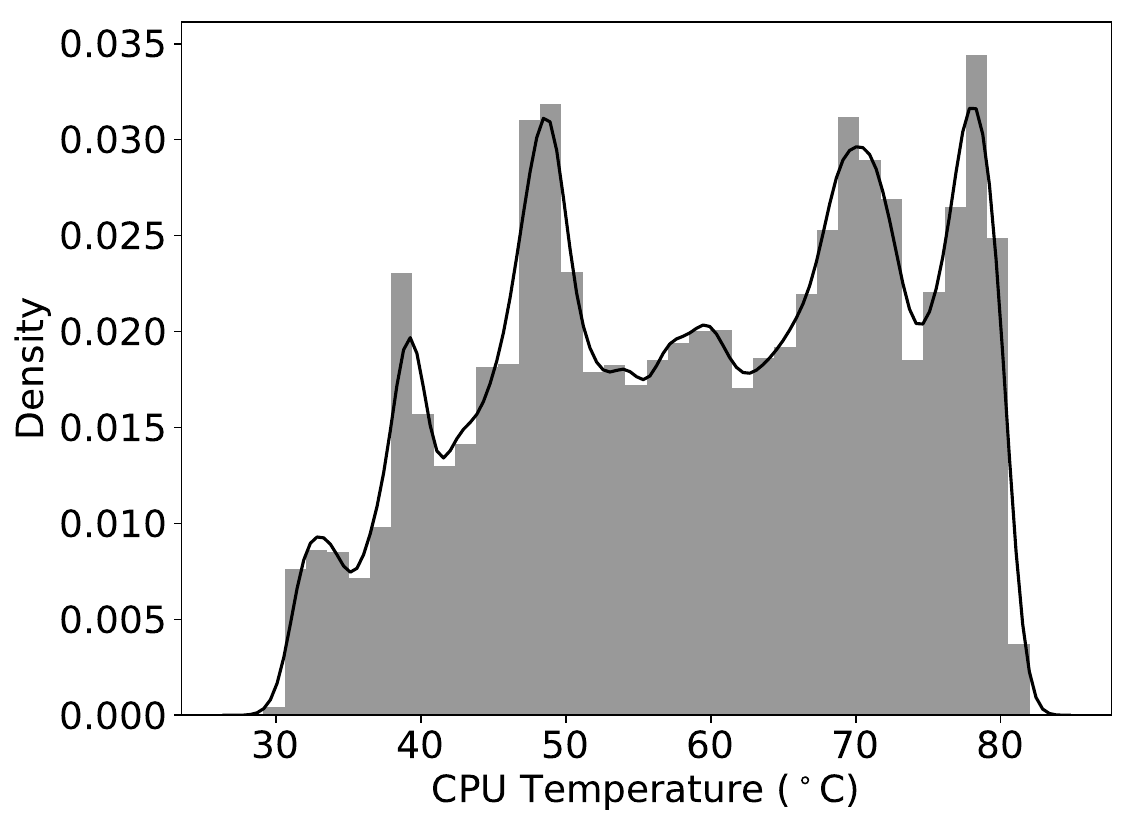}
            \caption{CPU temperature distribution}
            \label{cputemphistogram}
        \end{subfigure}
        \caption{ Feature set correlation  and temperature distribution}
        \label{motivation}
    \end{figure}
Thermal management is a  critical component in cloud data center operations. The presence of multi-tenant users and their heterogeneous workloads exhibit non-coherent behavior with respect to the thermal and power consumption of hosts in a cloud data center. Reducing even one degree of temperature in cooling saves millions of dollars over the year in large scale data centers \cite{gao2014machine}. In addition, most data centers and servers are already equipped with monitoring infrastructure, that has several sensors to read the workload, power, and thermal parameters. Using this data to predict the temperature is cost-effective and feasible.  Thereby, to analyze the complex relationships between different parameters that influence the host temperature, we collected data from a private cloud and studied it for intrinsic information. This data includes resource usage and sensor data of power, thermal, and fan speed readings of hosts. The detailed information about the data and collection method is described in Section \ref{datacollectionsection}.

The correlation between different parameters (Table \ref{featureTable}) and temperature distribution in the data center can be observed in  Figure \ref{allcor} and \ref{cputemphistogram}.  These figures are drawn from the data recorded on 75 hosts over a 90 days period. The logging interval was 10 minutes (i.e.,$75\times90\times24\times6$ records). The correlation plot in Figure \ref{allcor}  is based on the standard pairwise Pearson correlation coefficient represented as a heat map.  Here, the correlation value ranges from -1 to  1, where the value is close to 1  for highly correlated features, 0 for no correlation, and -1 for the negative correlation. For better illustration, the values are represented as color shades as shown in the figure. In addition, the correlation matrix is clustered based on pairwise Euclidean distance to enhance interpretability. It is evident that the CPU temperature of a host is highly influenced by power consumption and CPU load. However, factors like memory usage and machine fan speeds also have some degree of interdependence with it. Additionally, inlet temperature has a positive correlation with fan speeds and the number of VMs running on a host.

The high number of hosts operating at a peak  CPU temperature can be observed from Figure \ref{cputemphistogram}. The figure represents a histogram of the temperature distribution of all hosts. Thereby each bin on the $x$  axis represents a  quantized CPU temperature and the $y$ axis the corresponding probability density value. CPU temperature of hosts can reach more than 80 $\celsius$ and the occurrence of such conditions are numerous which is evidenced by high-density value on the $y$ axis for the respective bin.  In addition,  hosts exhibit inconsistent thermal behavior based on several factors. This non-linear behavior of hosts presents a severe challenge in temperature estimation. A single theoretical mathematical model,  applied even for homogeneous nodes, fails to accurately predict the temperature.  Two homogeneous nodes at a similar CPU load observe different CPU temperatures. For instance, at a CPU load of 50\% of the different hosts in our data set,  CPU temperature varies up to 14 $\celsius$.   Furthermore, with similar cooling settings, inlet temperature also varies up to  9 $\celsius$ between hosts.  These temperature variations are caused by factors like physical attributes such as the host's location, thermodynamic effects, heat recirculation, and thermal throttling mechanisms induced by the operating system based on workload behaviors \cite{ZhangTPDS2018}. Therefore, a temperature estimation model should consider the non-linear composite relationship between hosts.

Motivated by these factors,  we try to rely on data-driven prediction approaches compared to existing rigid analytical and expensive  CFD based methods.  We use the collected data to build the prediction models to accurately estimate the host temperature. Furthermore, guided by these prediction models, we propose a simple dynamic scheduling algorithm to minimize the peak temperature in the data center.

\section{System Model and Data-Driven Temperature  Prediction}
In this section, we describe the system model and discuss methods and approaches for cloud data center temperature prediction. We use these methods to further optimize our prediction model in Section \ref{sectionxgboost}.
\subsection{System Model}
\begin{figure}[t]
\centering
\includegraphics[width = 0.9\linewidth]{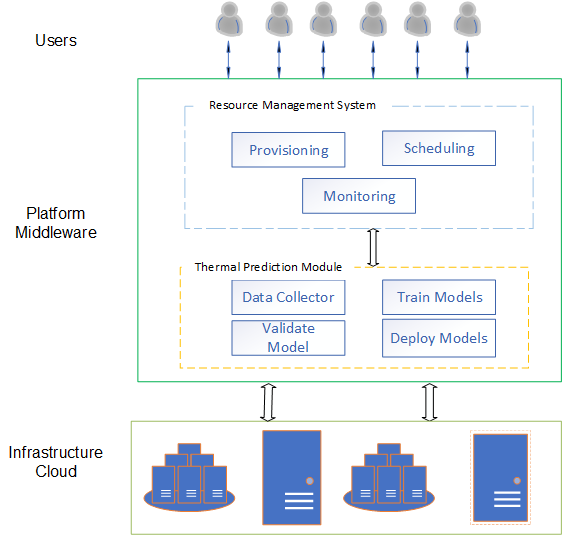}  %
\caption{System model}
\label{systemModel}
\end{figure}
A system model for predictive thermal management in the cloud data center is shown in Figure \ref{systemModel}.       
A Resource Management System (RMS) interacts with both, the users and the thermal prediction module, to efficiently manage the underlying resources of the cloud infrastructure. The prediction module consists of four main components, i.e., data collector, training the suitable model, validating the performance of the model, and finally deploying it for runtime usage.   RMS in a data center can use these deployed models to efficiently manage the resources and reduce the cost.  The important elements of the framework are discussed in the following subsections. 

\subsection{Data Collection}\label{datacollectionsection}  
    \begin{table}
    \centering
    \caption{Definition of features collected}
    \label{featureTable}
      \begin{tabular}{ll}
       \hline
           \textbf{Features} & \textbf{Definition} \\ \hline
            $CPU$                 &               CPU Load (\%)           \\           
            $R$     &             RAM- Random Access Memory  (MB) \\
            $R_x$      &       RAM  in usage (MB)                     \\ 
            $N_{CPU}$     &       Number of CPU cores                     \\
            $N_{CPUx}$     &     Number of CPU cores in use           \\
            $N_{Rx}$        &   Network inbound traffic (Kbps)         \\ 
            $N_{Tx}$     &  Network outbound traffic (Kbps)          \\ 
            $P_{c}$                     & Power consumed by host (watts)       \\
            $T_{cpu1}$                  &  CPU 1 temperature ($\celsius$)      \\ 
            $T_{cpu2}$                  &  CPU 2 temperature ($\celsius$)    \\ 
            $fs_1$     &   fan1 speed (RPM)    \\ 
            $fs_2$     &   fan2 speed  (RPM)     \\ 
            $fs_3$     &   fan3 speed    (RPM)  \\ 
            $fs_4$     &   fan4 speed    (RPM)  \\ 
            $T_{in}$           &  Inlet temperature   ($\celsius$)      \\ 
            $N_{vm}$           &   Number of VMs running on host                   \\ 
              \hline
      \end{tabular}
    \end{table}

\begin{table*}[]
 \centering
 \caption{Description of the  feature set variations in the dataset (aggregated from all the hosts)}
     \label{datadescription} 
\resizebox{\linewidth}{!}{    
\begin{tabular}{|l|l|l|l|l|l|l|l|l|l|l|l|l|l|l|}
\hline
     & \textbf{$CPU (\%)$} & \textbf{$R_x$} & \textbf{$N_{Rx}$} & \textbf{$N_{Tx}$} & \textbf{$N_{vm}$} & \textbf{$N_{CPUx}$} & \textbf{$P_{c}$} & \textbf{$fs_2$} & \textbf{$fs_1$} & \textbf{$fs_3$} & \textbf{$fs_4$} & \textbf{$T_{cpu1}$} & \textbf{$T_{cpu2}$} & \textbf{$T_{in}$} \\ \hline
Min  & 0                   & 3974           & 0                 & 0                 & 0                 & 0                   & 55.86            & 5636            & 5686            & 5688            & 5645            & 29.14               & 25.46               & 13.33             \\ \hline
Max  & 64.74               & 514614         & 583123.08         & 463888.76         & 21                & 101                 & 380.53           & 13469           & 13524           & 13468           & 13454           & 82                  & 75.96               & 18.05             \\ \hline
Mean & 18.09               & 307384.48      & 2849.00           & 1354.164          & 9                 & 54                  & 222.73           & 9484            & 9501            & 9490            & 9480            & 59.50               & 50.78               & 25.75             \\ \hline
\end{tabular}
}
\end{table*}

\begin{table}[]
\caption{Private cloud data collected for this work }
     \label{datacollection}
\resizebox{\columnwidth}{!}{ 
\begin{tabular}{|c|c|c|c|c|c|}
\hline
\textbf{\#Hosts} & \textbf{\#VMs} & \textbf{\begin{tabular}[c]{@{}c@{}}Total CPU\\ Cores\end{tabular}} & \textbf{\begin{tabular}[c]{@{}c@{}}Total\\ Memory\end{tabular}} & \textbf{\begin{tabular}[c]{@{}c@{}}Collection \\ Period\end{tabular}} & \textbf{\begin{tabular}[c]{@{}c@{}}Collection \\ Interval\end{tabular}} \\ \hline
75               & 650            & 9600                                                               & 38692 GB                                                        & 90 days                                                               & 10 Minute                                                               \\ \hline
\end{tabular}
}
\end{table}

 An ML-based prediction model is as good as the data it has been used to train. In the data center domain, training data can include application and physical level features to train the model \cite{ZhangTPDS2018}. The application features include instruction count,  number of CPU cycles, cache metrics (read, write and miss), etc. Accordingly, physical features include host-level resource usage (CPU, RAM, I/O, etc.) and several sensor readings (power, CPU temperature, fan speeds). Relying on both of these features is feasible in bare metal HPC data centers where administrators have exclusive access to the application and physical features. However, in the case of Infrastructure as Service (IaaS) clouds,  resources are virtualized and provisioned as VMs or containers, thus, giving users exclusive isolated access to the application execution environment. The presence of a hypervisor or container-based virtualization in IaaS clouds restricts access to application-specific features. Moreover, a diverse set of users in the cloud have a different type of workloads exhibiting different application behaviors which impede cloud RMS to rely on application-specific features. As a consequence, to predict host temperature, the RMS is required to monitor fine-grained resource usage and physical features of the host system that can be directly accessed. In this regard, we show that this data is adequate to predict the host temperature accurately.

The Melbourne Research Cloud (MRC)\textcolor{black}{ \footnote{https://docs.cloud.unimelb.edu.au/}} provides Virtual Machines (VM) to students and researchers. The representative data is collected from a subset of machines from  MRC. This computing infrastructure provides computing facilities to students and researchers as a virtual machine (VM). We collect data from a subset of the total machines in this cloud.  A brief summary of this data is presented in Table \ref{datacollection}. It includes logs of  75 physical hosts having an average number of 650  VMs. The data is recorded for a period of  3 months and the log interval is set to  10 minutes. The total count of resources includes 9600 CPU cores and 38692 GB of memory.  After data filtration and cleaning, the final dataset contains 984712 tuples, each host approximately having around  13000 tuples. Each tuple contains 16 features including resource and usage metrics, power, thermal, and fan speed sensors measurements. The details of these features are given in Table \ref{featureTable}.  As each host is equipped with two distinct  CPUs,  two temperature measurements are reported per machine. In addition, each system has four separate fans installed to provide cooling. \textcolor{black}{ The reason to collect data for an extended period is to capture all the dynamics and variations of parameters to train the model effectively. This is only possible when host resources have experienced different usage levels over time. A  model built over such data allows accurate prediction in dynamic workload conditions.  An overview of variations of all parameters is depicted in Table \ref{datadescription} ( $N_{CPU}$ and $R$ are not included as they represent constant resource capacity)}. 

To collect this data, we run a  collectd\footnote{https://collectd.org/} daemon on every host in the data center, which is a standard open-source application that collects system and application performance counters periodically through system interfaces such as IPMI and sensors.  These metrics are accessed through network API's and stored in a centralized server in the  CSV format. We used several bash and python scripts to pre-process the data. Specifically, python pandas\footnote{https://pandas.pydata.org/} package to clean and sanitize the data. All invalid measurements (e.g. $NaN$) were removed.  For the broader use of this data to the research community and for the sake of reproducibility, we will publish the data and scripts used in this work.

\subsection{Prediction Algorithms}
The choice of regression-based algorithms for our problem is natural since we aim to estimate the numerical output variable i.e.,  temperature. In the search for suitable prediction mechanisms, we have explored different ML algorithms including different regression techniques, such as Linear Regression (LR), Bayesian Regression (BR), Lasso Linear Regression (LLR), Stochastic Gradient Descent regression (SGD), an Artificial  Neural Network (ANN) model called Multilayer Perceptron (MLP), and an ensemble learning technique called gradient boosting, specifically, eXtreme Gradient Boosting (XGBoost). 

\textcolor{black}{Since each host in our cluster has two CPUs that are jointly controlled by the same operating system (which may dynamically move workloads between them), we always regard the maximum of the respective two CPU temperature measurements as the systems' effective CPU temperature.} We aim to build a model for each host to accurately capture its thermal behavior properties. For that reason, instead of solely predicting CPU temperature,   we predict the host ambient temperature ($T_{amb}$)  which is a combination of inlet temperature and CPU temperature  \cite{cool}. 
 The reason to consider ambient temperature instead of  CPU temperature is manifold. First, by combining the inlet and CPU temperature, it is feasible to capture thermal variations that are induced by both the inlet and CPU temperature (cause of these variations are discussed in Section \ref{sec:motivation}). Second, at a data center level, cooling settings knobs are adjusted based on host ambient temperature rather than individual CPU temperature \cite{Moore2006weatherman}. In addition, resource management systems in the data center consider host-level ambient temperature as a threshold parameter whereas operating system level resource management techniques rely on CPU temperature.

Therefore, to build the prediction model for individual hosts, we parse the data set and partition it based on host IDs. For each individual host,  the feature set consists of a variable number of tuples, with each tuple having these features ($CPU$, $R$, $R_x$, $N_{CPU}$,  $N_{CPUx}$, $N_{Rx}$, $N_{Tx}$, $N_{vm}$,  $P_c$, $fs_1-fs_4$, $T_{amb}$). Note that, we have excluded inlet and CPU temperatures from the list, as we have combined these as ambient temperature ($T_{amb}$) which is our target prediction variable. 

We used sci-kit learn package  \cite{pedregosa2011scikit} to implement all the algorithms. For XGBoost, we used a standard python package\footnote{https://github.com/dmlc/xgboost}  available on Github.  
\textcolor{black}{The parameters for each of the algorithms are set to their default settings in our implementation}. For MLP, it follows a standard  3 layers architecture, with the number of neurons at a hidden layer set to 5 and a single output neuron, and  'ReLu' as the activation function. 

To avoid the overfitting of the models,  we adopt k-fold cross-validation where the value of k is set to 10. Furthermore, to evaluate the goodness of fit for different models,  we use the  Root Mean Square Error (RMSE) metric which is a  standard evaluation metric in regression-based problems \cite{caruana2006empirical}. The RMSE is defined as follows. 

\begin{equation}
 RMSE = \sqrt{\frac{1}{n}\Sigma_{i=1}^{n}{\Big({y_i -\hat{y_i}}\Big)^2}}
\label{rmseequation}
\end{equation}
In Equation \ref{rmseequation}, $y_i$ is  the observed value, $\hat{y_i}$ is the predicted output variable, and $n$ is the total number of predictions. 
The value of RMSE represents the standard deviation of the residuals or prediction errors. 
\textcolor{black}{The prediction models attempt to minimize an expectation of loss, thus, lower RMSE values are preferred.} 

\begin{figure}
\centering
   \includegraphics[width=1\linewidth]{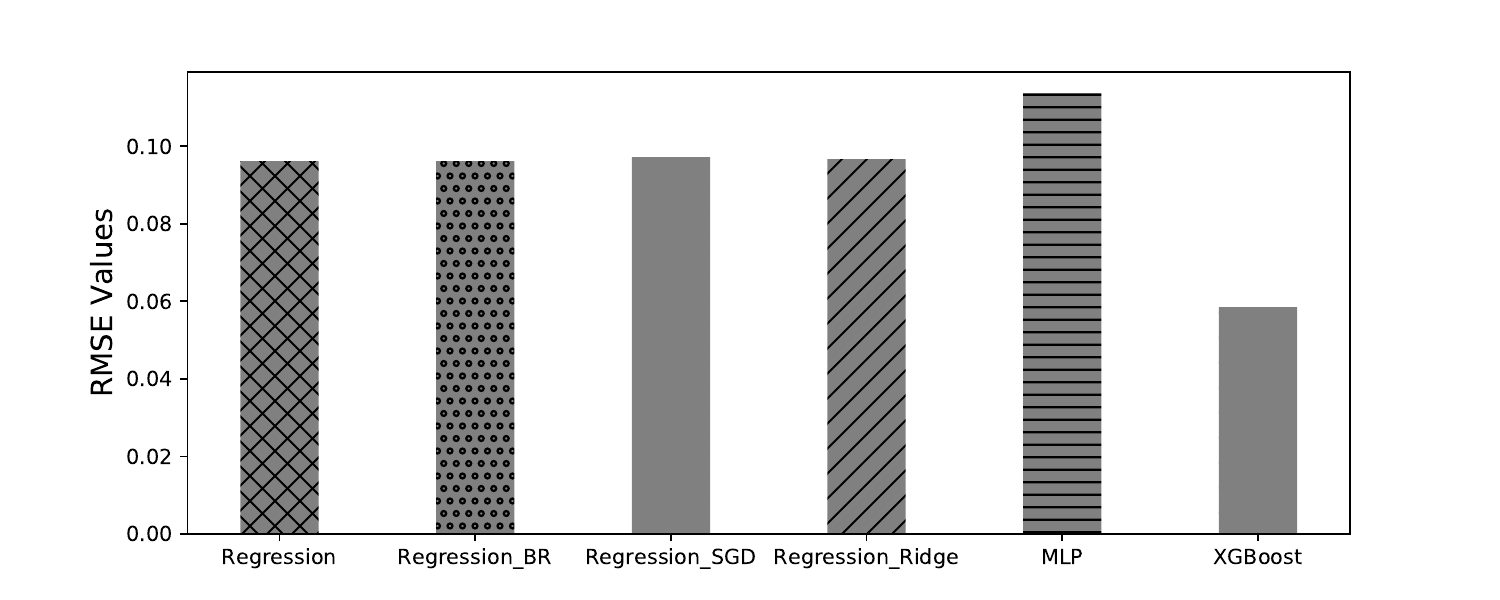}
   \label{rmsecomparison}
\caption{Average prediction error between different models}
\label{rmseresult}
\end{figure}

The performance of different algorithms is shown in Figure \ref{rmseresult}. These results are an average of all the hosts' prediction model results. In  Figure \ref{rmseresult}, we can observe that XGBoost has a very low  RMSE value, indicating that, the residuals or prediction errors are less and its predictions are more accurate.  We observed that MLP has a high error value compared to other algorithms.
In addition, different regression variants have performed almost similar to each other.   As the gradient boosting method  XGBoost results are promising, we focus more on this algorithm to explore it further, optimize, and adapt it for further scheduling as explained in Section \ref{scheduling}.

\section{Learning with Extreme Gradient Boosting}\label{sectionxgboost}
Boosting is an ensemble-based machine learning method that builds strong learners based on weak learners. Gradient boosting is an ensemble of weak learners, usually decision trees.   XGBoost (eXtreme Gradient Boosting) is a    scalable, fast and efficient gradient boosting variant for tree boosting proposed by Chen et al \cite{chen2016xgboost}. \textcolor{black}{  It incorporates many advanced techniques to increase the performance, such as parallelism, cache optimization with better data structure, and out of core computation using block compression and block sharing techniques which is essential  to  prevent the memory overflow in training large data sets on constrained resource environments.} \textcolor{black}{Accordingly, the impact of boosting techniques including XGBoost is evidenced by its dominant adoption in many  Kaggle competitions  and also in large scale production systems \cite{bekkerman2015present, mangal2016using, bianchini2020toward}.}

\begin{figure*}
\centering
  \captionsetup{justification=centering}
\begin{subfigure}[t]{0.45\textwidth}
 \raisebox{-\height}{\includegraphics[width=1\linewidth]{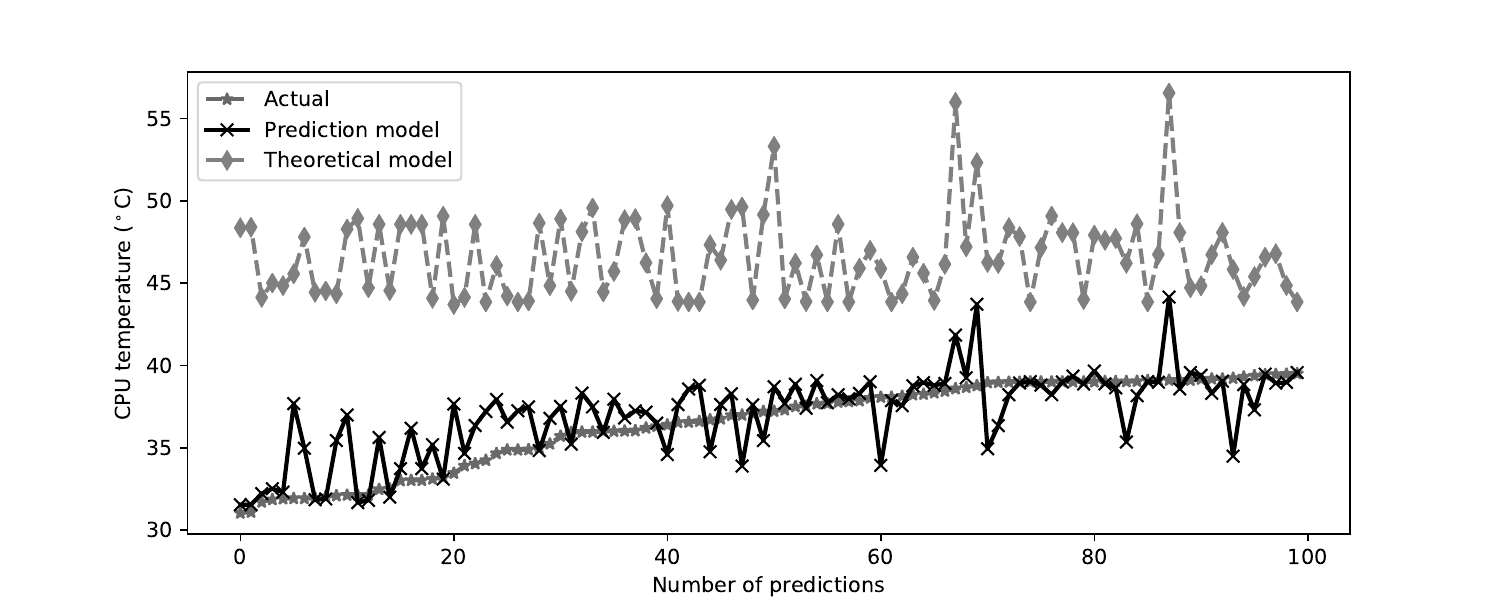}} %
   \caption{Temperature estimation compared to actual values}
   \label{theortocalmodelcomparison}
\end{subfigure}
\begin{subfigure}[t]{0.45\textwidth}
 \raisebox{-\height}{\includegraphics[width=1\linewidth]{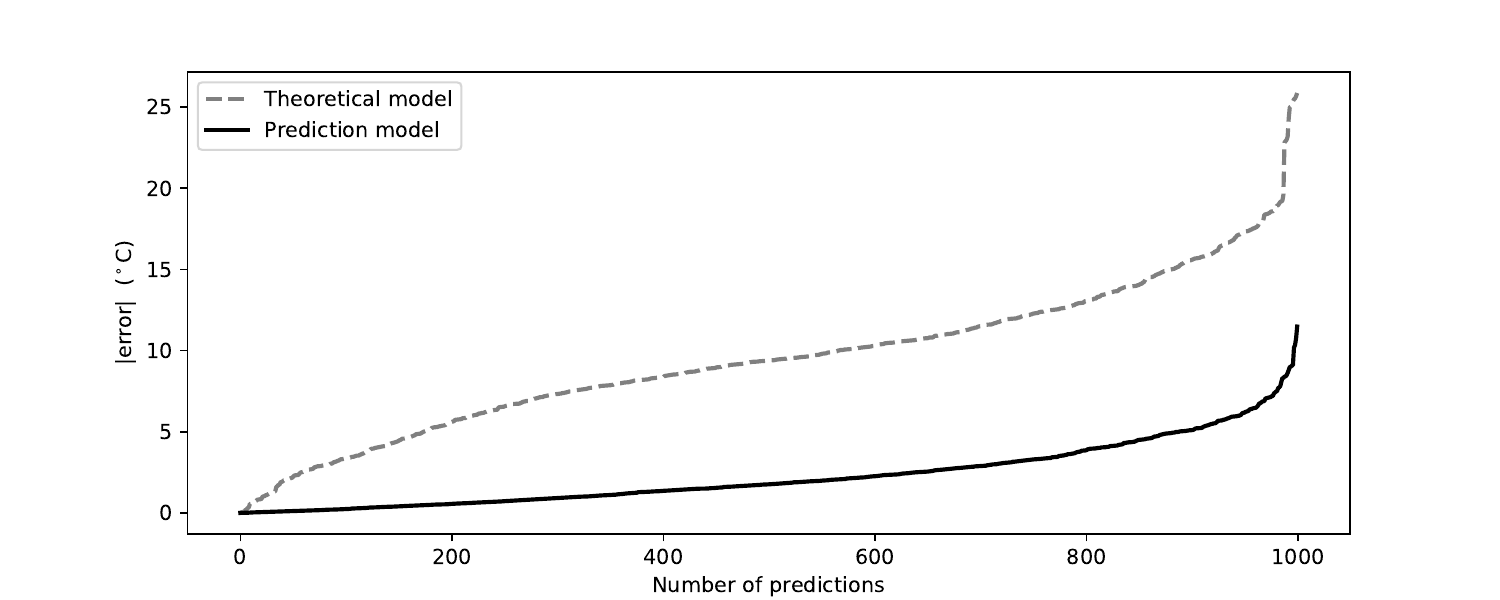}}
   \caption{Rank order of prediction errors}
   \label{rankordererror}
\end{subfigure}
\caption{Comparison of  prediction   and theoretical model}
   \label{theoreticalmodelsresult}
\end{figure*}

The XGBoost algorithm is an ensemble of K Classification or Regression Trees (CART) \cite{chen2016xgboost}. This can be used for both classification and regression purpose.  The model is trained by using an additive strategy. For a dataset with n instances and m features, the ensemble model uses k additive functions to estimate the output. Here,  $x$ being a set of input features, $x = \lbrace x_1, x_2, ... x_m\rbrace$ and $y$ is the target prediction variable.
\begin{equation}
\hat{y_{i}}  = \phi\left ( x_i \right ) = \sum_{k=1}^{K}{f_{k}\left ( x_i \right )} ,  \hspace{4mm}    f_k \in  F
\label{ensemble}
\end{equation}

In the   Equation \ref{ensemble}, $F$ is space  of  all the regression trees, i.e,  $F = \lbrace f\left(x\right)  = w_{q\left(x\right)}\rbrace $, and $\left( q: \mathbb{R}^m \rightarrow T, w \in \mathbb{R}^T \right)$. Here, q is the structure of each tree which maps to corresponding leaf index. T represents total number of leaves in the tree. each  $f_k$ represents an independent tree with structure $q$ and leaf weights $w$. To learn the set of  functions used  in the model, XGBoost minimizes the following regularized objective.

\begin{multline}
\zeta\left( \phi\right)  = \sum_{i}{l\left (\hat{y_{i}}, y_i \right )} + \sum_{k}{\Omega\left( f_{k}\right)},  \\ 
where  \quad  \Omega\left( f\right) = \gamma T = 1\frac2\lambda\parallel w \parallel^2
\label{objectivexgboost}
\end{multline}

In   Equation \ref{objectivexgboost}, the first term $l$ is the differentiable convex loss function that calculates the difference between  predicted value $\hat{y_i}$, observed value $y_i$. \textcolor{black}{ $\Omega$ penalizes the complexity of the model to control overfitting. Thereby, T  is the number of nodes in the tree and $w$ is assigned values for each leaf node of the tree.
This regularized objective function attempts to select a model based on simple predictive functions.} %

We use the grid search technique to find the optimal parameters to further enhance the performance of the model. \textcolor{black}{Here, the $\gamma$ parameter is used to decide the minimum loss reduction required to make a further partition on a leaf node of the tree. Subsample ratio decides the amount of sampling selected from training data to grow the trees.} Accordingly, the optimal values for  $\gamma$ are 0.5, the learning rate is 0.1,  maximum depth of the tree is 4,  minimum child weight is 4, and the subsample ratio is 1, and rest of the parameters are set to default. With these settings, the best RMSE value achieved is 0.05. It is important to note that the prediction based temperature estimation is feasible for any data center given the historical data collected from the individual data center.

\section{Evaluating the Prediction Model with Theoretical Model}

To evaluate the feasibility of our temperature prediction models, we compare the prediction results to extensively used theoretical analytical model \cite{Zhang2007cputemp, tang2008energy, sun2017spatio}. Here, the temperature estimation is based on the RC model which is formulated from  analytical methods.  The temperature of a host ($T$) is calculated based on the following equation.
\begin{equation}
    T = PR + T_{in} + (T_{initial} -PR - T_{in})\times e^{-{\frac{t}{RC}}}
    \label{cputemp}
\end{equation}
In  Equation \ref{cputemp},  $P$ is the dynamic power of host, $R$ and $C$ are thermal resistance ($k/w$) and heat capacity ($j/k$) of the host respectively. $T_{initial}$ is the initial temperature of the CPU. Since analytical models estimate CPU temperature, we also predict CPU temperature to compare the results instead of ambient temperature. 

To compare the results, we randomly select 1000 tuples from our whole dataset and analyze the result between prediction and theoretical models. For the theoretical model, the value of $P$ and $T_{in}$ are directly used from our test data set. The value of thermal resistance ($R$) and heat capacity ($C$) is set as 0.34 $K/w$ and 340 $J/K$ respectively and  $T_{initital}$  is set to 318 $K$ \cite{Zhang2007cputemp}.

The performance of the two models in temperature estimation can be observed in Figure \ref{theoreticalmodelsresult}. For the sake of visibility, Figure \ref{theortocalmodelcomparison}  includes 100 tuples of data. As the figure suggests,  our proposed model based on XGBoost's estimation is very close to the actual values, whereas the theoretical model has a large variation from the actual values. Figure \ref{rankordererror}, represents a rank order of the absolute errors (from actual temperature) of two models in  $\celsius$.  The theoretical model deviates as far as 25 $\celsius$ from the actual values. In this test, the average error of the theoretical model is 9.33 $\celsius$ and our prediction model is 2.38 $\celsius$. These results reflect the feasibility of using prediction models over theoretical models for temperature estimation. \textcolor{black}{It is important to note that, the prediction models need to be trained for different data centers separately with well-calibrated data that have enough data points to cover all temperature and load conditions in order to predict temperature accurately.} Nevertheless, in the absence of such a facility, it is still feasible to use theoretical analytical models that rely on a minimum number of simple parameters.

\section{Dynamic Scheduling guided by Prediction Models}\label{scheduling}
Applications of temperature predictions are numerous. It can be used to change the cooling settings such as supply air temperature to save the cooling cost \cite{cool}. It is also useful in identifying the thermal anomalies which increase the risk of failures and injects performance bottlenecks. Moreover, one foremost usage would be in a data center resource management system's tasks such as resource provisioning and scheduling. 

With the given historical host's data, predictive models are trained and deployed for runtime inference. A scheduling algorithm invokes a deployed prediction model to accurately predict the host temperature. The input to the prediction model is a set of host features. In our model,  the features can be easily collected from the host's onboard sensors. These features are accessed from the host's system interface through HTTP APIs. The complexity to retrieve this input feature set information is  $O(1)$.  The latency of this operation depends on the data center's local network capabilities. Moreover,  the models need to be retrained only when changes are introduced to the data center environment,  like, the addition of new hosts or change in the physical location of hosts. Considering the fact that such changes are not so frequent in a data center, the cost of building and using such predictive models in resource management tasks like scheduling is highly feasible.

 In this regard, we propose dynamic scheduling of VMs in a cloud data center based on the temperature prediction model we have proposed. Here, we intend to reduce the peak temperature of the system while consolidating VMs on fewest hosts as possible for each scheduling interval which is a preferred way to reduce the energy in a cloud data center \cite{beloglazov2012energy}.   In this problem,  $n$ physical hosts in data center hosting $m$ VMs at timestep $t$, the objective is to reduce the number of active hosts in a data center at $t+1$ by consolidating the VMs based on workload level. This consolidation process inside the data center is critical and carried out regularly to reduce overall data center energy \cite{verma2009server, brownout}.  This procedure mainly includes three steps. First, identifying under loaded hosts from which we can potentially migrate VMs and shut down the machine. Also finding overloaded hosts and migrate VMs from them to reduce the risk of Service Level Agreements (SLA) violation, here, SLA is providing requested resources to VMs without degrading their performance. Second, selecting VMs for migration from the over-and underloaded hosts identified in previous step, and finally, identifying new target hosts to schedule the selected VMs. The scheduling for consolidation process  allows hosts to experience high load and potentially reach the threshold temperature which is useful in evaluating our prediction models effectively. Therefore, The objective of our problem is defined as follows:

    \begin{mini}|l|
  {   }{T^{peak} 
 =     \sum_{t=0}^{T} \sum_{j=1}^{m}\sum_{i=1}^{n} \delta_{ji}^{t}T_{i}^{t}}{}{}
 \label{optimalequation}
 \addConstraint{u(h_{i}) \leq U_{max} } 
\addConstraint{T_{i}^{t}} {< T_{red}} 
\addConstraint{ \sum_{j=0}^{m}VM_{ji}(R_{cpu}, R_{mem}) \leq  h_{i}(R_{cpu}, R_{mem})}
 \addConstraint{\delta_{ji}^{t}  =} {\lbrace 0,1 \rbrace }                                            
 \addConstraint{\sum_{i=1}^{n} \delta_{ji}^{t}} { = 1}
    \end{mini}
    
The objective function in  Equation \ref{optimalequation}  minimizes the peak temperature of the hosts while scheduling VMs dynamically in all the time steps $t = \lbrace 0,... \quad  T\rbrace$. Here, list of VMs that are to be scheduled are represented with the index $j$ where $j =\lbrace1,... \quad m\rbrace$, and list of candidate hosts as  $i$, where $i = \lbrace1, ...\quad n\rbrace$. The $T_{i}^{t}$ indicates  temperature of host $i$ at time $t$. The constraints ensure that potential thermal and CPU thresholds are not violated due to increased workload allocation.  They also assure the capacity constraints, i.e, a host is considered as suitable only if enough resources are available for VM ($R_{cpu}, R_{mem}$). Here, $\delta_{ji}^{t}$ is a binary with the value  1 if the $VM_{j}$ is allocated to $host_i$ at time interval $t$, otherwise, 0. The summation of $\delta_{ji}^{t}$ is equal to 1, indicating that $VM_{j}$ is allocated to at most 1 host at time $t$. The objective function in  Equation \ref{optimalequation} is executed at each scheduling interval to decide the target host for the VMs to be migrated. Finding an optimal solution for the above equation is an NP-hard problem and it is infeasible for on-line dynamic scheduling. Accordingly, to achieve the stated objective and provide a near-optimal approximate solution within a reasonable amount of time, we propose a simple Thermal-Aware heuristic Scheduling (TAS) algorithm that minimizes the peak temperature of data center hosts.
 
To dynamically consolidate the workloads (VMs) based on current usage level, our proposed greedy heuristic scheduling algorithm \ref{mainalgo} is executed for every scheduling interval.  The input to the algorithm is a list of VMs that are needed to schedule. These are identified based on overload and underload condition. To identify overloaded hosts, we use CPU ($U_{max}$) and temperature threshold ($T_{red}$) together.  In addition, if all the VMs from a host can be migrated to current active hosts, the host is considered as an underloaded host. The VMs that are to be migrated from overloaded hosts are selected based on their minimum migration time, which is the ratio between their memory usage and available bandwidth \cite{beloglazov2012energy}. The output is scheduling maps representing target hosts for those VMs.  For each VM to be migrated (line 2), Algorithm \ref{mainalgo} tries to allocate a new target host from the active list. In this process, algorithm initializes necessary objects (lines 3-5) and the prediction model is invoked to predict the accurate temperature of a host (line 7). The VM is allocated to a host that has the lowest temperature among active hosts (lines 8-11). This ensures the reduction of peak temperature in the data center and also avoids potential hotspots resulting in lower cooling cost. Moreover, this algorithm also assures the constraints listed in Equation \ref{optimalequation} are met (line 10), so that added workload will not create a potential hotspot by violating threshold temperature ($T_{red}$). In addition, resource requirements of VM ( VM($R_x$)) are satisfied, and the CPU utilization threshold is within the limit ($U_{max}$). If no suitable host is found in the process, a new idle or inactive host is allocated (line 16) from the available resource pool.
\begin{algorithm}[]
\begin{flushleft}
\textbf{Input: }$VMList$- List of VMs to be scheduled\\
\textbf{Output: } Scheduling Maps
\end{flushleft}
\caption{Thermal Aware Dynamic Scheduling to Minimize Peak Temperature}
\label{mainalgo}
\begin{algorithmic}[1]
\FOR{\texttt{t $\leftarrow$ 0 to T}}
  \FORALL { $vm$ in $VMList$}{
          \STATE $allocatedHost$ $\leftarrow$  $\emptyset$
          \STATE $hostList$ $\leftarrow$  {Get list of active hosts}
            \STATE $minTemperature$ $\leftarrow$  $maxValue$  
              \FORALL { $host$ in $hostList$ }{
                 \STATE $\hat{T_i}$ $\leftarrow$ {Predict temperature by invoking prediction model}
                \IF  {($\hat{T_i}$ $<$ $minTemperature$) }  
                        \STATE$minTemperature$ $\leftarrow$   $\hat{T_i}$         
                        \IF  {($\hat{T_i}$ $<$ $T_{red}$  and $u(h_{i}) \leq U_{max}$ and $vm(R_x)$ $<$ $host(R_x)$) }
                           \STATE $allocatedHost$ $\leftarrow$ $host$
                        \ENDIF 
                    \ENDIF 
            } %
      \ENDFOR %
    \IF{allocatedHost $==$ $\emptyset$}
        \STATE allocatedHost $\leftarrow$ Get a new host from inactive hosts list %
    \ENDIF
}%
\ENDFOR %
\ENDFOR %
\end{algorithmic}
\end{algorithm}

The algorithm \ref{mainalgo} has a worst-case complexity of  $\mathcal{O}(VN)$, which is a polynomial-time complexity. Here,  $\mid V \mid$ is the number of VMs to be migrated, and $\mid N \mid$ is a number of hosts in a data center. 

\section{Performance Evaluation}
In this section, we evaluate the performance of the proposed algorithm coupled with our prediction model and compare and analyze the results with baseline algorithms.
\subsection{Experimental Setup}
We evaluated the proposed thermal aware dynamic scheduling algorithm through CloudSim toolkit \cite{calheiros2011cloudsim}. We extended CloudSim to incorporate the thermal elements and implement algorithm \ref{mainalgo}. We used a real-world dataset from  Bitbrain \cite{shen2015statisticalBitBrain}, which has traces of resource consumption metrics of business-critical workload hosted on Bitbrain's infrastructure. This data includes logs of over  1000 VMs workloads hosted on two types of machines. We have chosen this data set as it represents real-world cloud Infrastructure usage patterns and the metrics in this data set are similar to the features we have collected in our data set (Table \ref{featureTable}). This is useful to construct precise input vectors for prediction models. 

The total experiment period is set to 24 hours and the scheduling interval to 10 minutes, which is similar to our data collection interval. Note that, in the algorithm, prediction models are invoked in many places. The prediction is required to identify the host with the lowest temperature, to determine a host overloaded condition,  and also to ensure thermal constraints by predicting their future time step temperature.

To depict the experiments in a real-world setting, we model host configurations similar to the hosts in our data center, i.e.,  DELL C6320 machines.  This machine has an Intel Xeon E5-2600 processor with  dual CPUs (32 cores each)  and 512 GB RAM.  \textcolor{black}{The VMs are configured based on  the  VM flavours in our research cloud \textcolor{black}{\footnote{https://docs.cloud.unimelb.edu.au/guides/allocations/}}. We choose four VM types from general flavors,  configuration of these VMs are presented in Table \ref{hostandvmtypes}}.  The number of hosts in the data center configuration is 75, similar to the number of hosts in our private cloud collected data, and the number of VMs is set to 750, which is the maximum number possible on these hosts based on their maximum resource requirements. The workload is generated to these VMs according to  Bitbrain's dataset.

The  CPU threshold ($U_{max}$) is set to  0.9. According to the American Society of Heating, Refrigerating and Air-Conditioning Engineers (ASHRAE) \cite{ASHRAE} guidelines, the safe operable temperature threshold for data center hosts is in-between 95 to 105 $\celsius$. This threshold is a combined value of CPU temperature and inlet temperature together. Accordingly we set temperature threshold ($T_{red}$) to 105 $\celsius$.

The new target machines for VMs to be scheduled are found based on algorithm \ref{mainalgo}. This requires predicting the temperature of hosts in the data center. If the  $host_i$   temperature is predicted ($\hat{T_{i}}$) at the beginning of timestep $t+1$ then the input to prediction model is a single vector consisting of a set of features ($CPU$, $P_c$, $fs_1-fs_4$,  $N_{CPU}$, $N_{CPUx}$, $R$, $R_x$, $N_{Rx}$, $N_{Tx}$, $N_{vm}$) representing its resource and usage metrics along with the power and fan speed measurements. The resource usage metrics are easily gathered from host utilization levels based on its currently hosted VMs' workload level. To estimate the power $\hat{P_{i}} $, we use SPECpower benchmark \cite{spec}, which provides accurate power consumption (in watts) for our modeled host (DELL C6320)  based on  CPU utilization. We estimate fan speeds from simple regression using remaining features to simplify the problem.

We export the trained models as serialized python objects and expose them to our scheduling algorithm by hosting on HTTP Flask application \footnote{http://flask.pocoo.org}. The CloudSim scheduling entities invoke the prediction model through REST APIs by giving feature vector and host ID as input,  the HTTP application returns predicted temperature for the associated host.

{\color{black}
  \begin{table}[]
        \centering
        \caption{\textcolor {black}{VM Configurations}}
        \label{hostandvmtypes}
           \begin{tabular}{lllll}
             \hline

               \textcolor {black}{ Name} & \textcolor {black}{Core}  & \textcolor {black}{RAM}  \\ 
                                \hline

                \textcolor {black}{VM1 (uom.general.1c4g)}   & \textcolor {black}{1}    & \textcolor {black}{4 GB }  \\
               \textcolor {black}{VM2 (uom.general.2c8g)}   & \textcolor {black}{2}     & \textcolor {black}{8 GB}  \\
                \textcolor {black}{VM3 (uom.general.4c16g)}  & \textcolor {black}{4}    & \textcolor {black}{16 GB} \\
                \textcolor {black}{VM4 (uom.general.8c32g)}  & \textcolor {black}{8}          & \textcolor {black}{32 GB} \\
                \hline
            \end{tabular}
    \end{table}

}
\subsection{Analysis of Results}
\begin{figure}
\centering
   \includegraphics[width=1\columnwidth]{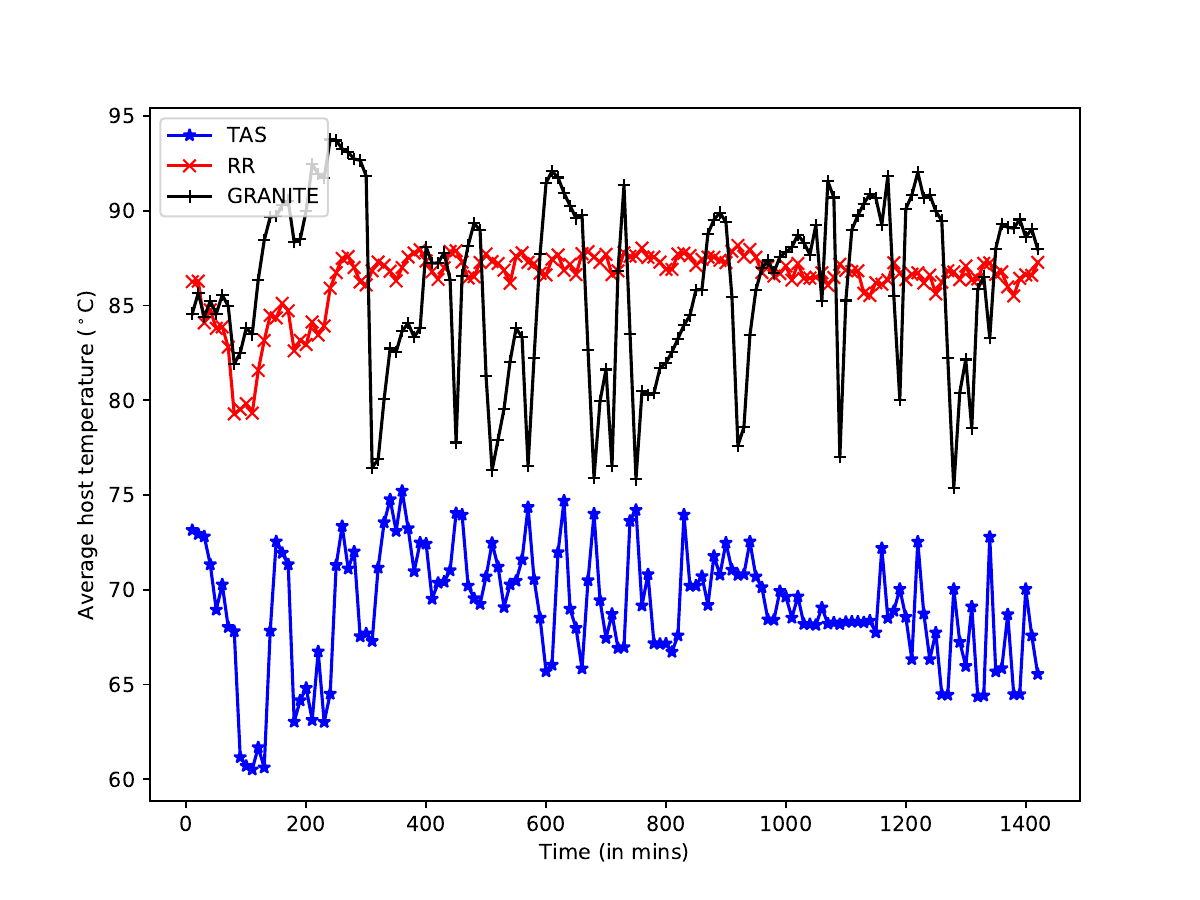} %
   \caption{Average temperature in each scheduling interval (total experiment time of 24 hours, with scheduling interval of 10 minute)}
   \label{averagetemperature}
\end{figure}

\begin{figure*}
  \captionsetup{justification=centering}
\begin{subfigure}[t]{0.32\textwidth}
\includegraphics[width=\linewidth]{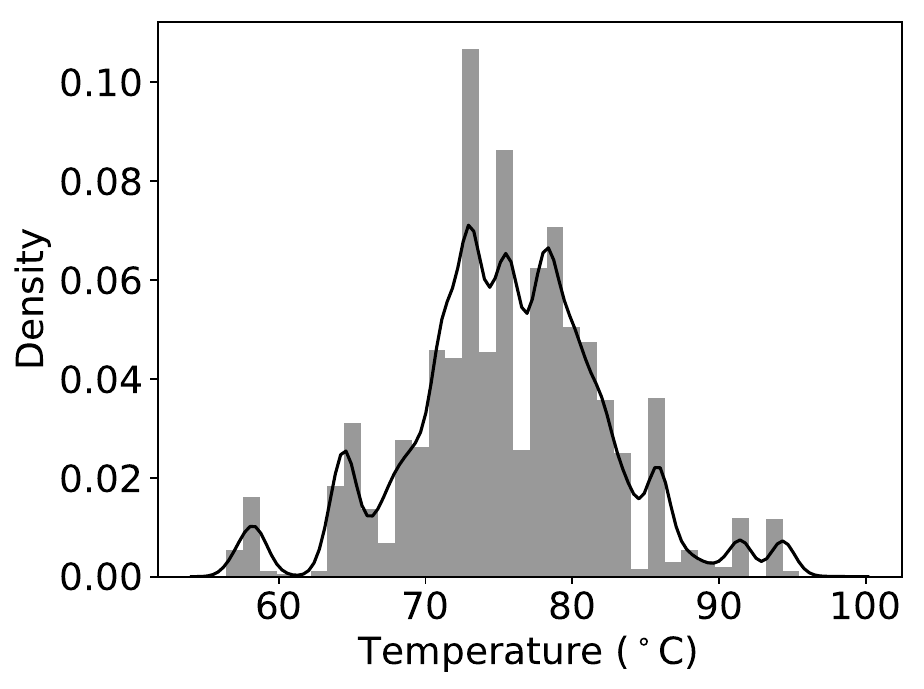} %
   \caption{TAS}
   \label{TAShistogram}
\end{subfigure}
\begin{subfigure}[t]{0.32\textwidth}
\includegraphics[width=\linewidth]{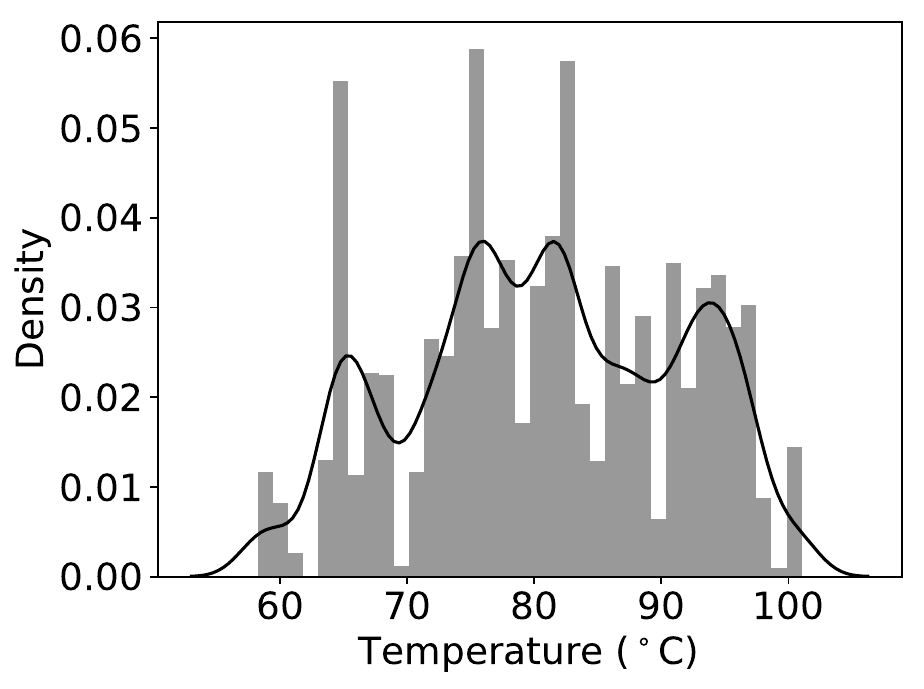}
   \caption{RR}
   \label{RRhistogram}
\end{subfigure}
\begin{subfigure}[t]{0.32\textwidth}
\includegraphics[width=\linewidth]{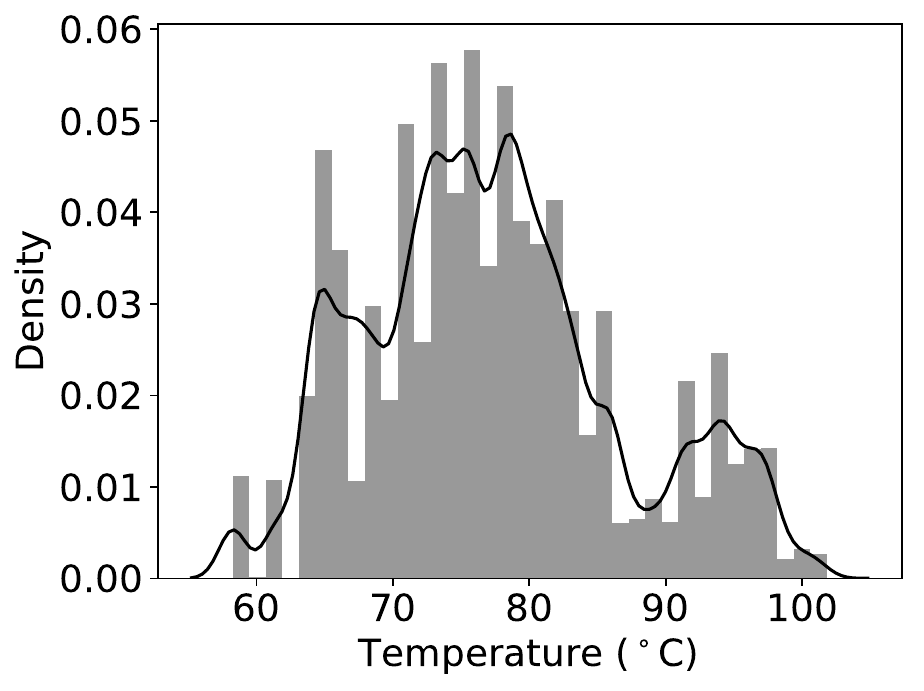}
   \caption{GRANITE}
   \label{Granitehistogram}
\end{subfigure}
\caption{Temperature distribution analysis due to scheduling (aggregated from all hosts in experimented period)}
   \label{schedulinghistogram}
\end{figure*}

\begin{figure}[]{}
   \includegraphics[width=1\linewidth]{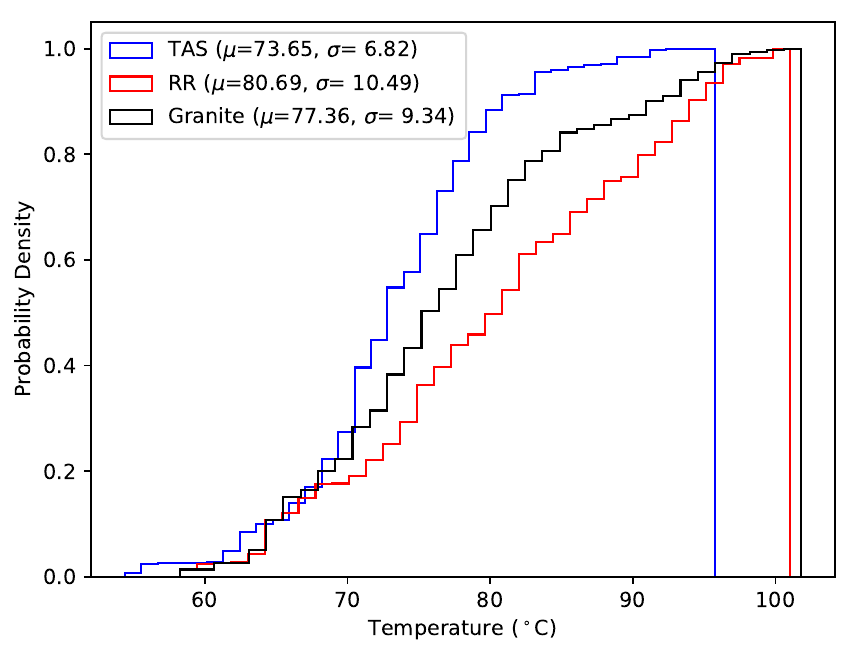}
   \caption{\textcolor{black}{CDF between TAS and RR and GRANITE}}
   \label{CDF}
\end{figure}

\begin{figure*}
\centering
  \captionsetup{justification=centering}
\begin{subfigure}[t]{0.45\textwidth}
 \raisebox{-\height}{\includegraphics[width=\linewidth]{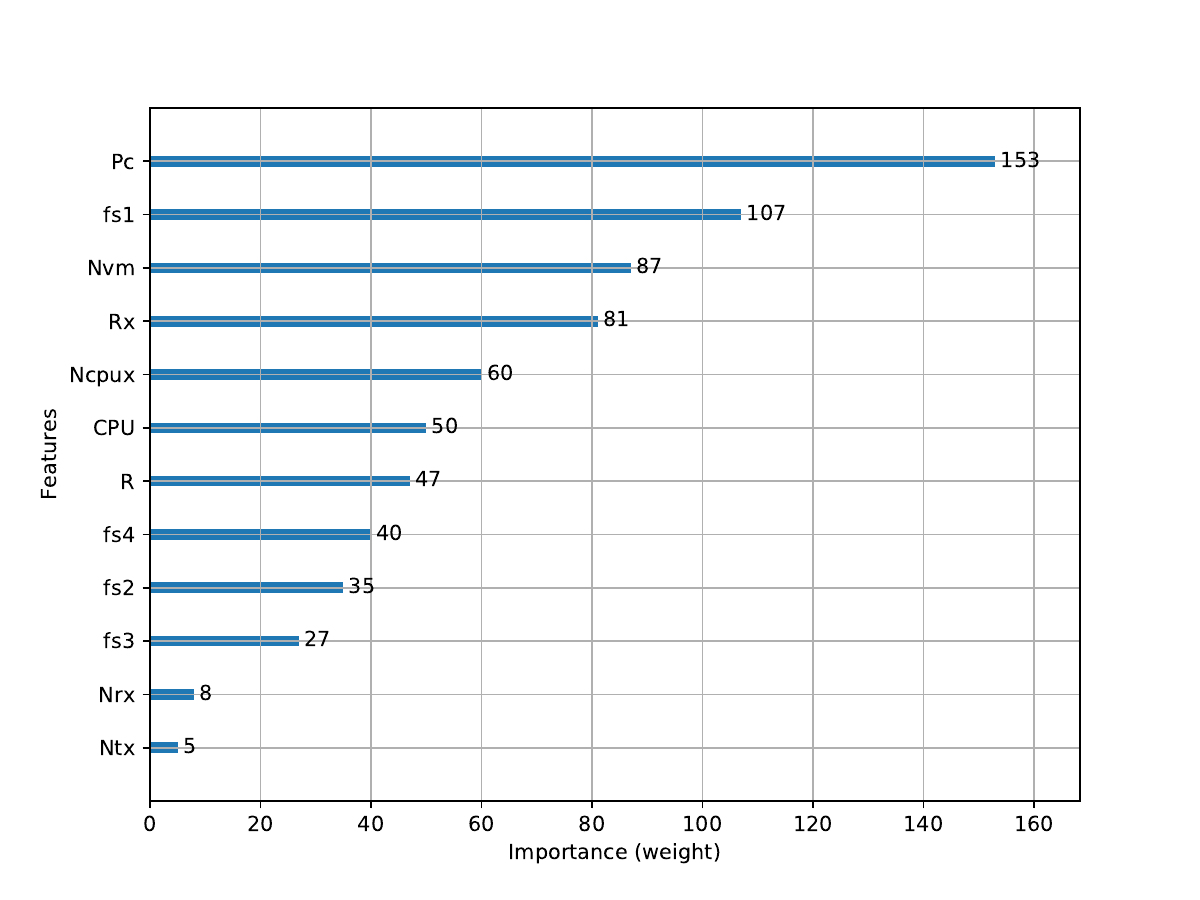} }
   \caption{Feature importance (weight)- number of times a particular feature occurs in the trees}
   \label{featureimportance}
\end{subfigure}
\begin{subfigure}[t]{0.45\textwidth}
 \raisebox{-\height}{\includegraphics[width=\linewidth]{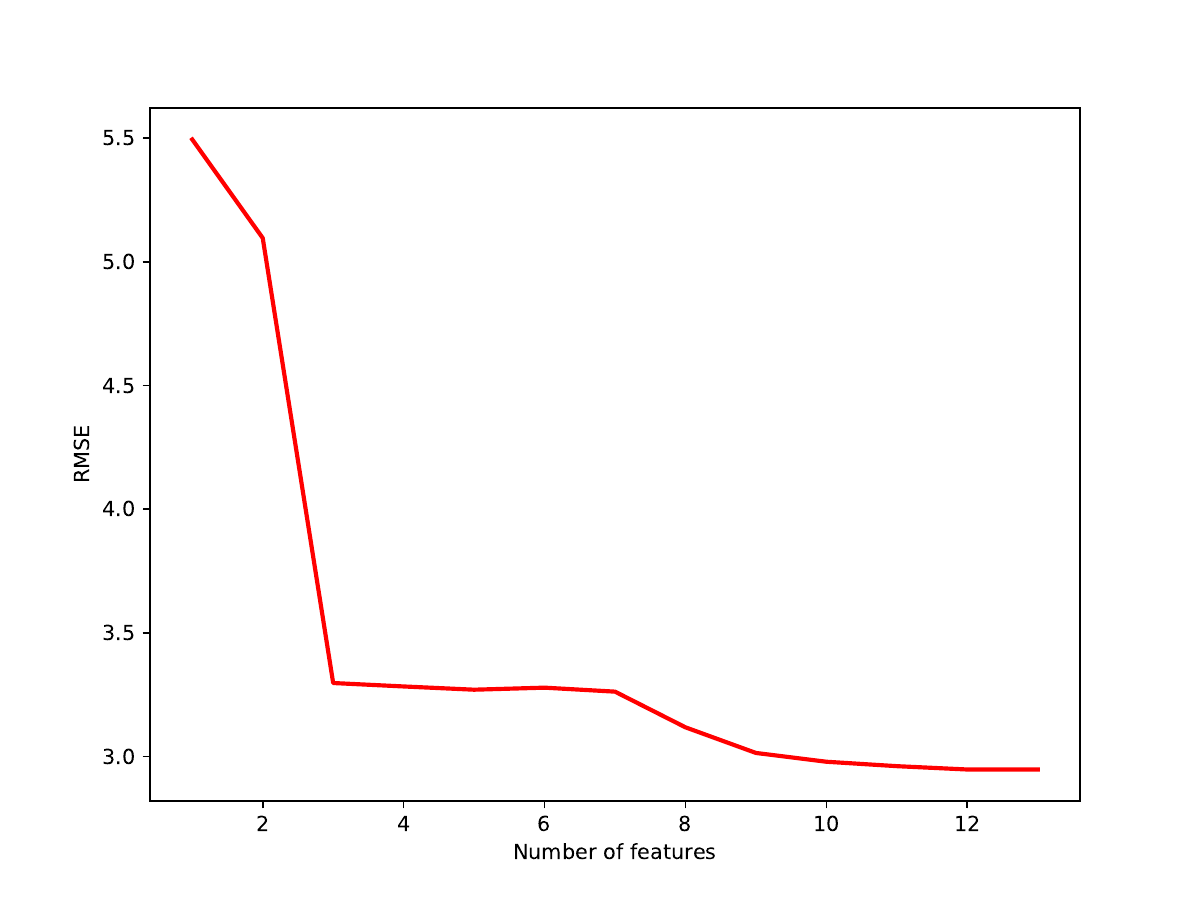}}
    \caption{Feature threshold analysis}
   \label{featurethreshold}
\end{subfigure}

\caption{Feature analysis}
   \label{thresholdlmodelsresult}
\end{figure*}

We compare the results with two baseline algorithms as shown below.
\begin{itemize}
\item Round Robin (RR) -  This algorithm tries to distribute the workload equally among all hosts by placing  VMs on hosts in a circular fashion.  The similar constraints are applied as in algorithm \ref{mainalgo}. We show that the notion of equal distribution of workloads fails to minimize the peak temperature and thermal variations in a data center.
\item GRANITE- This is a thermal-aware    VM scheduling algorithm proposed in \cite{holistictpds} that minimizes computing and cooling energy holistically. We choose this particular algorithm, because, similar to us,  it also addresses the thermal-aware dynamic VM scheduling problem.  
 
\end{itemize}   

\textcolor{black}{We use our prediction models to estimate the temperature in both RR and GRANITE algorithms. For GRANITE, the required  parameters are set similar to their algorithm in \cite{holistictpds} including overload and underload detection methods.}
The comparison of the average temperature from all hosts in each scheduling interval by all three algorithms is shown in Figure \ref{averagetemperature}. Our   Thermal-Aware Scheduling (TAS) has the lowest average temperature compared to RR and GRANITE. The RR algorithms' equal workload distribution policy results in less variation in average temperature. However, this will not help to reduce the peak temperature in the data center irrespective of its intuitive equal distribution behavior as it doesn't consider the thermal behavior of individual hosts and its decisions are completely thermal agnostic. The GRANITE policy has a high average temperature and large variations between scheduling intervals due to its inherent dynamic threshold policies.  To further analyze the distribution of temperature due to two scheduling approaches, we draw a histogram with Kernel Density Estimation (KDE)  by collecting temperature data from all the hosts in each scheduling interval as shown in Figure \ref{schedulinghistogram}. Most of the hosts in the data center operate around 70 to 80 $\celsius$ in TAS (Figure \ref{TAShistogram}), well below the threshold due to its expected peak temperature minimizing objective. However, the RR approach results in more thermal variations with sustained high temperatures (Figure \ref{RRhistogram}). The GRANITE also has significant distributions around the peak temperature (\ref{Granitehistogram}).  This temperature distribution is effectively summarized using the Cumulative Distribution Function (CDF) between three approaches (Figure \ref{CDF}). As we can see in Figure \ref{CDF}, TAS reaches the probability density value of  1  well below 100 $\celsius$, indicating most of the hosts operate in reduced temperature value.  RR and GRANITE has a peak temperature of more than 100 $\celsius$ with high cumulative probability. \textcolor{black}{ In addition, as depicted in Figure \ref{CDF}, the average  and standard deviation of temperature in TAS ($\mu = 75.65$, $\sigma = 6.82$) is lesser compared to the  other two approaches ($\mu = 80.69$, $\sigma = 10.49$ for RR and $\mu = 77.36$, $\sigma = 9.34$ for Granite ), this is also evidenced by Figure \ref{averagetemperature}.}   

Further results of the experiments are depicted in Table \ref{schedulingresult}. The total energy consumption by TAS, RR, and GRANITE is 172.20 kWh,   391.57 kWh, and  263.20 kWh, respectively (the total energy is a combination of cooling and computing energy calculated as in \cite{etas}). Therefore, RR and GRANITE have 56 \% and 34.5 \% more energy consumption than  TAS, respectively.  This is because RR and GRANITE distribute workload into more hosts resulting in a high number of active hosts. In this experimented period, RR and  GRANITE had 18 and 11 average number of active hosts while the TAS algorithm resulted in 4 active hosts. Furthermore, although RR distributes workload among many hosts, its thermal agnostic nature had a  peak temperature of 101.44 $\celsius$, GRANITE had peak temperature of 101.80 $\celsius$   and  TAS had attained a maximum of 95.5 $\celsius$ during the experimentation period which is 6.5 $\celsius$ lower than the latter approaches. This demonstrates that the accurate prediction of host temperature with an effective scheduling strategy can reduce the peak temperature and also save a significant amount of energy in the data center.

\begin{table}[]
\caption{Scheduling results compared with RR and GRANITE algorithm }
     \label{schedulingresult}
      \resizebox{0.9\columnwidth}{!}{    
\begin{tabular}{|c|c|c|c|}
\hline
\textbf{Algorithm} & \textbf{\begin{tabular}[c]{@{}c@{}}Peak Temperature\\   ( $\celsius$)\end{tabular}} & \textbf{\begin{tabular}[c]{@{}c@{}}Total Energy\\   (kwh)\end{tabular}} & \textbf{\begin{tabular}[c]{@{}c@{}}Active \\ Hosts\end{tabular}} \\ \hline
TAS                & 95                                                                           & 172.20                                                                  & 4                                                                \\ \hline
RR                 & 101.44                                                                       & 391.57                                                                  & 18                                                               \\ \hline
GRANITE            & 101.81                                                                       & 263.20                                                                  & 11                                                               \\ \hline
\end{tabular}
}
\end{table}
{\color{black}
\subsection{Evaluating Performance Overhead}\label{sec:perf-overhead}
\begin{figure*}
\centering
  \captionsetup{justification=centering}
\begin{subfigure}[t]{0.24\textwidth}
 \raisebox{-\height}{\includegraphics[width=\linewidth]{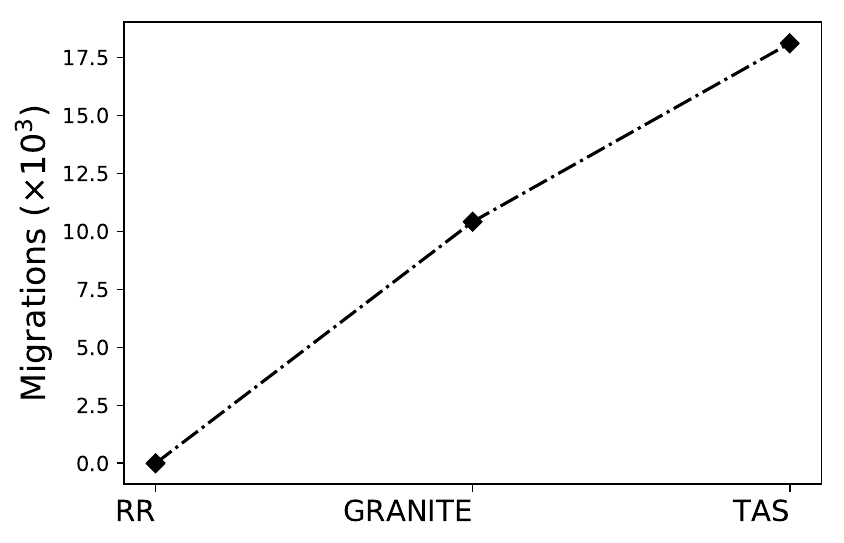} }
   \caption{ \textcolor{black}{Number of VM migrations}}
   \label{mighrations}
\end{subfigure}
\begin{subfigure}[t]{0.24\textwidth}
 \raisebox{-\height}{\includegraphics[width=\linewidth]{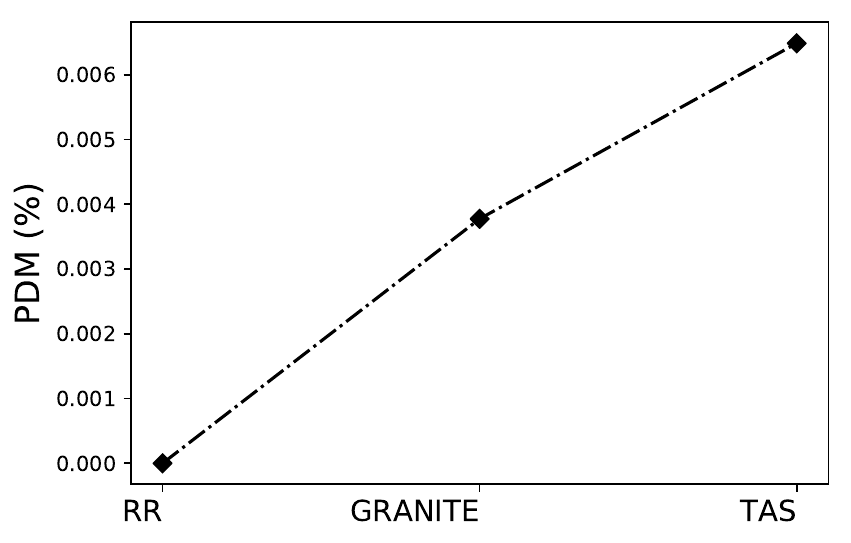}}
    \caption{\textcolor{black}{The PDM metric}}
   \label{pdm}
\end{subfigure}
\begin{subfigure}[t]{0.24\textwidth}
 \raisebox{-\height}{\includegraphics[width=\linewidth]{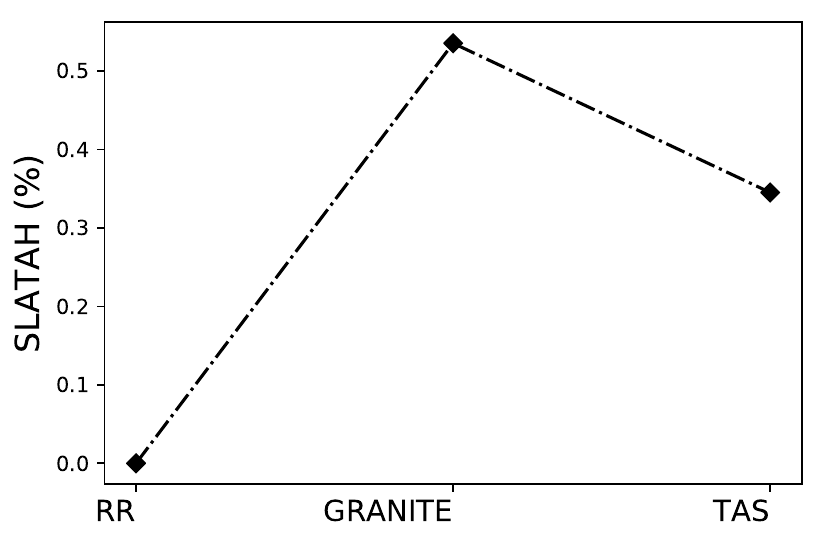}}
    \caption{\textcolor{black}{The SLATAH metric}}
   \label{slatah}
\end{subfigure}
\begin{subfigure}[t]{0.24\textwidth}
 \raisebox{-\height}{\includegraphics[width=\linewidth]{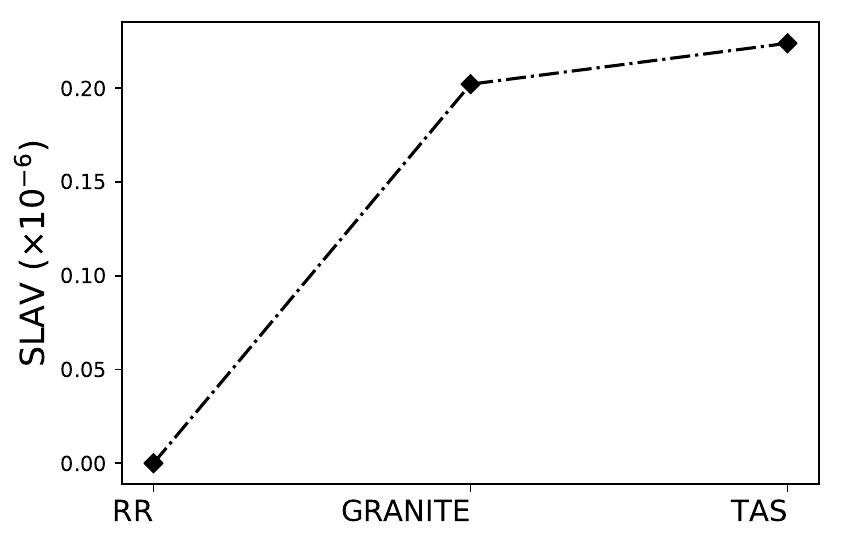}}
    \caption{\textcolor{black}{The $SLA_{violation}$ metric}}
   \label{slav}
\end{subfigure}
\caption{\textcolor{black}{Performance Overhead Metrics}}
   \label{slametrics}
\end{figure*}

It is important to estimate the overhead of dynamic scheduling caused due to migration and workload consolidation.  In the context of scheduling in the cloud, the expected performance is usually defined using Service Level Agreements (SLAs). In our approach, the scheduling is at a higher VM level, hence, we represent the  SLA  metrics using the  VM level features. In this regard,  we consider the following metrics  \cite{holistictpds, beloglazov2012energy}: \\
\textbf{Number of VM migrations:}   Virtual machines may experience degraded performance during migration. Hence, the number of migrations should be minimized to reduce the overhead and avoid SLA violations. \\
\textbf{SLA\textsubscript{violation}:} 
Due to oversubscription and consolidation,  hosts may reach full utilization level (100\%), in such cases, the VMs on such host experiences degraded performance. This is expressed using  SLA violation Time per Active Host ($SLA_{TAH}$) metric as shown in Equation \ref{slatah}. Furthermore, the consolidation of VMs comes with performance overhead caused due to live VM migration \cite{voorsluys2009cost}, this  Performance Degradation due to Migration (PDM) is defined as in Equation \ref{PDMEqn}.

\begin{align}
                SLA_{TAH} =  \frac{1}{  N} \sum_{i=1}^{N}\frac{ T_{max}} {T_{active}}
                \label{SLATAH}
    \end{align}
\begin{align}
                PDM =  \frac{1}{ M} \sum_{j=1}^{ M}\frac{ {C_{A{j}} - C_{R{j}}}} {C_{R{j}}}
                \label{PDMEqn}
\end{align}

\begin{align}
                SLA_{violation} =  SLA_{TAH}\times PDM
                \label{overallSLA}
\end{align}%
Here, $N$ is total number of hosts, $T_{max}$ is amount of time $Host_i$ has experienced 100\% of utilization and $T_{active}$ is total active time of $Host_{i}$.  $M$ is the total number of VMs. The  $C_{A{j}}$ is the total amount of CPU capacity allocated and $C_{R{j}}$ is the total amount of CPU capacity requested by $VM_j$ while in migration during its lifetime, this captures the under allocation of VMs during live migration. The overall SLA violation of cloud infrastructure ($SLA_{violation}$) can be defined by combining both  $SLA_{TAH}$ and $PDM$  metrics as shown in Equation  \ref{overallSLA}. 

The results of overhead metrics for different algorithms are shown in Figure \ref{slametrics}. As shown in Figure \ref{mighrations}, the number of migrations is 10417 and 18117 for GRANITE and  TAS, respectively. The RR has zero migrations. It is expected as RR distributes workload equally among the required number of hosts from the initial step and is not concerned about dynamic optimizations in runtime. For the  $PDM$ metric (Figure \ref{pdm}), GRANITE and TAS have 0.0037 \% and 0.0064\%, respectively. This is because to TAS has a higher number of migrations compared to GRANITE. As TAS continuously tries to minimize the peak temperature among active hosts based on workload level, it performs aggressive consolidation in each scheduling interval. However, the proactive approach of TAS trying to reduce the host peak of temperature also results in reduced CPU  overload of hosts. This is evidenced as the TAS has a lower value of  $SLA_{TAH}$ metric (0.34\%) compared to the GRANITE (0.53\%).  Furthermore, for the  overall $SLA_{violation}$ metric (Figure \ref{slav}), TAS has  increased value  ($0.22 \times 10^{-6}$) compared to  GRANITE ( $0.20\times10^{-6}$). This little increased value is due to the higher $PDM$ value of TAS. However, TAS significantly outperforms both GRANITE and RR  in reducing peak temperature and energy efficiency with this negligible overhead. 

} %
\subsection{Dealing with False Predictions}
 In our scheduling experiments, we observed that a few of the temperature predictions have resulted in some large number which is beyond the boundaries of the expected value.  A further close study into such cases has revealed that this happens with particularly three hosts which were almost idle in the data collection period of 3 months having a CPU load less than 1\%, which means the models trained for these hosts have limited variations in their feature set.  As the trained models did not have any instance close to the instance of prediction,  prediction results in an extreme variant value. Such a false prediction in runtime results in an incorrect scheduling decision that affects the normal behavior of the system. In this regard, the scheduling process should consider such adverse edge cases. To tackle this problem, we set minimum and maximum bound for expected prediction value based on our observations in the dataset. For any prediction beyond these boundaries, we pass the input vector to all remaining hosts' models and take an average of predicted value as a final prediction value. In this way, we try to avoid the bias influenced by a particular host and also get a reasonably good prediction result. In the case of a huge number of hosts, subsets of hosts can be used for this. 

\textcolor{black}{This also suggests that,  to effectively use the prediction models,  the training data should have a  distribution of values of all hosts covering all possible ranges of values. Deploying such models in a real-world data center requires good coverage of data to handle all possible operating points    of the data center so that when  ML models are trained they will not be overfitted for a skewed range of data and thus perform poorly}.

{\color{black}
\subsection{Assumptions and Applicability}
The scheduling algorithm and prediction models proposed in this paper have the following assumptions and applicabilities. The scheduling algorithm is applicable for workloads that run in VMs for a long period without any interruptions (such as web and enterprise applications). Our policy tries to monitor the utilisation level of such workloads and consolidate them at regular intervals for energy efficiency while minimising the data center's peak temperature. The workload independent performance metrics  in section \ref{sec:perf-overhead}  indirectly captures the overhead of the scheduling algorithm.  For other types of workloads such as tasks with predefined completion time, this algorithm is not directly applicable. In addition, the models trained from the particular data center should only be used in that  data center. This is required to capture the unique characteristics and configuration of a data center that influences temperature variations in it. They include data center physical rack-layout, air circulation pattern, and server heat dissipation rate that directly affects the sensor readings and thus ambient temperature of server \cite{tang2008energy, Zhang2015minimize, holistictpds}. Hence, it is essential to train  prediction models with data collected from a individual data center to capture its characteristics. However, our proposed techniques are still applicable in building such models. Therefore,  the scheduling algorithm and prediction models are only suitable for a specific workloads,  in a particular data center.
}
 \section{Feature Set Analysis}

We carried out a feature analysis to identify the importance of each feature towards the model performance. This analysis can also be used in the feature selection process to remove the redundant features,   reduce the computational cost, and increase the performance. Figure \ref{featureimportance} shows the importance of each feature in the constructed XGBoost model. Here, the weight metric associated with each feature corresponds to its respective number of occurrences in the constructed tree which indirectly notifies its importance.  Based on the results, host power ($P_c$), fanspeed1 ($fs_1$) and number of VMs ($N_vm$) are  the most important features towards accurate prediction. It is important to note that, though we have 4 fan speeds, the model intuitively selects one fan speed with more weight, this is since all four fans operate almost at the same rpm, which is observed in our data set. The least important feature is network metrics ($N_{rx}$, $N_{tx}$) along with the remaining three fan speed readings. The crucial observation is that the model gives high importance to power instead of CPU load, indicating, the high correlation between temperature and power. The number of cores ($NC$) is not included in the tree as it has constant value across hosts introducing no variation in the data.

The performance of temperature prediction with different thresholds can be observed in Figure \ref{featurethreshold}. We start with the most important feature and recursively add more features according to their importance to the model. The $y$ axis indicates RMSE value and the $x$ axis shows a number of features. The first three features ($P_c$,$fs_1$,$N_{vm}$) significantly contribute to prediction accuracy and the accuracy gain is little as we add more features to the model. Therefore, based on the required accuracy or RMSE value, we can select top $n$ features to effectively train the model with less complexity. 

\section{Related Work}   
Thermal management using theoretical analytical models has been studied by many researchers in the recent past \cite{cool, cao2017cooling, sun2017spatio, tang2008energy}. These models based on mathematical relationships to estimate the temperature are not accurate enough when compared to the actual values. \textcolor{black}{Moreover, \cite{cao2017cooling, tang2008energy} uses analytical models and targets HPC systems where jobs have specific completion time, while our work target the virtualized cloud datacenters with long-running applications that need dynamic scheduling and migration in realtime.} Furthermore, some of the studies have also explored using  CFD models \cite{choi2008cfd}.  Computational Fluid Dynamics (CFD) models provide an accurate thermal measurement, however, their massive computational demand hinders their adoption in realtime online tasks such as scheduling. Researchers are audaciously exploring data-driven ML algorithms to optimize the computing system efficiency \cite{JeffMLforSystem, Fox2019}.  With the help of ML techniques, Google data centers are able to reduce up to 40 \% of their cooling costs \cite{gao2014machine}.

Many researchers in recent years study thermal and energy management inside the data center using machine learning techniques. The vast applications have been used for finding an optimal setting or configurations of systems to achieve energy efficiency \cite{ImesICPP2018}. However, ML techniques specific to temperature prediction are studied by   Zhang et al. \cite{Zhang2015minimize} where they proposed the Gaussian process-based host temperature prediction model in HPC data centers. They used a two-node Intel Xeon Phi cluster to run the HPC test applications and collect the training data. In addition, they also proposed a greedy algorithm for application placement to minimize the thermal variations across the system. In an extended work \cite{ZhangTPDS2018}, they enhanced their solution to include more efficient models such as lasso linear and Multilayer  Perceptron (MLP). The results have shown that predictive models are accurate and perform well in data center resource management aspects.  Imes et al. \cite{ImesICPP2018} explored different ML classifiers to configure the different hardware counters to achieve energy efficiency for a given application. They tested 15 different classifiers including Support Vector Machine (SVM), K-Nearest Neighbours (KNN), and Random Forest (RF), etc. This work only considers energy as an optimization metric ignoring the thermal aspect. Moreover, these works are specific to HPC data centers where temperature estimation is done for application-specific which requires access to application counters. Nevertheless,  our proposed solution is for Infrastructure clouds, where such an approach is not feasible due to limited access to application counters enforced by the isolated virtualized environment. Thus, we rely on features that completely surpass application counters and only consider host-level resource usage and hardware counters and yet achieve a  high prediction accuracy.

Furthermore, Ignacio et al.  \cite{SelfMapsAransay2015} showed the thermal anomaly detection technique using Artificial Neural Networks (ANNs). They specifically use  Self Organising Maps (SOM) to detect abnormal behavior in the data center from a previously trained reliable performance.  They evaluated their solution using traces of anomalies from a real data center. Moore et al. \cite{Moore2006weatherman} proposed Weatherman, a predictive thermal mapping framework for data centers. They studied the effect of workload distribution on cooling settings and temperature in the data center. These models are designed to find the thermal anomalies and manage the workload at a data center level without giving any attention to accurate temperature prediction.

In addition to thermal management, many others applied  ML techniques for scheduling in distributed systems to optimize the parameters such as energy, performance, and cost. Among many existing ML approaches,   Reinforcement Learning (RL) is widely used for this purpose \cite{Comput2018JPDCRL, DRLcloudCheng2018RL,  Mao2016RMSDRL}.  Orheab et al. \cite{Comput2018JPDCRL} studied the RL approach for scheduling in distributed systems. They used the Q-learning algorithm to train the model that learns optimal scheduling configurations. In addition, they proposed a platform that provides scheduling as a service for better execution time and efficiency. Cheng et al. proposed the DRL cloud, which provides an RL framework for provisioning and task scheduling in the cloud to increase energy efficiency and reduce the task execution time. Similarly, \cite{Mao2016RMSDRL} et al. studied deep RL based resource management in distributed systems. Learning to schedule is prominent with RL based methods due to the fact that RL models keep improving in runtime \cite{sutton1998introductionRL} which is convenient for scheduling. However, our work is different from these works in a way that, the primary objective of our problem is to estimate the data center host temperature accurately to facilitate the resource management system tasks. In this regard,  our work acts as complementary to these solutions where such thermal prediction models can be adopted by these ML-based scheduling frameworks to further enhance their efficiency.

 \section{Conclusions and Future Work}
Estimating the temperature in the data center is a complex and non-trivial problem.  Existing approaches for temperature prediction are inaccurate and computationally expensive. Optimal thermal management with accurate temperature prediction can reduce the operational cost of a data center and increase reliability.  Data-driven temperature estimation of hosts in a  data center can give us a more accurate prediction than simple mathematical models as we were able to take into consideration  CPU and inlet airflow temperature variations through measurements.   Our study which is based on physical host-level data collected from our University's private cloud has shown a large thermal variation present between hosts including CPU and inlet temperature. To accurately predict the host temperature, we explored several machine learning algorithms. Based on the results, we found a gradient boosting based XGBoost model for temperature prediction is the best. Our extensive empirical evaluation has achieved high prediction accuracy with the average RMSE value of 0.05. In other words, our prediction model has an average error of 2.38 $\celsius$. Compared to an existing theoretical model, it reduces the prediction error of 7 $\celsius$.

Guided by these prediction models, we proposed a dynamic scheduling algorithm for cloud workloads to minimize the peak temperature. The proposed algorithm is able to save up to 34.5\% more of energy and reduce up to 6.5 $\celsius$ of average peak temperature compared to the best baseline algorithm.  
\textcolor{black}{ It is important to note that, though the models built for one data center are optimized for its own (as each data center's physical environment and parameters vastly change),  the methodology presented in this work is generic and can be applied to any cloud data center given the sufficient amount of data collected from the respective data centers.}
    
    In the future, we plan to explore more sophisticated models to achieve better accuracy and performance. We also intend to extend the work for heterogeneous nodes like GPUs or FPGAs. Another interesting direction is to consider parameters related to weather and predictions and their effect on cooling and scheduling long jobs.
\section*{Acknowledgments}
We thank Bernard Meade and Justin Mammarella at Research Platform Services, The University of Melbourne for their support and
providing access to the infrastructure cloud and data. \textcolor{black}{This work is partially supported by a Discovery Project research grant funded by the ARC (the Australian Research Council).}

\bibliographystyle{IEEEtran}

\bibliography{references}

\begin{IEEEbiography}
[{\includegraphics[width=1in,height=1in,clip,keepaspectratio]{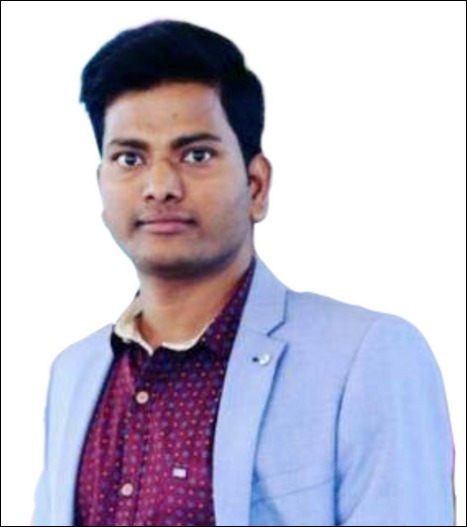}}]
{Shashikant Ilager}
is a PhD candidate with the Cloud Computing and Distributed Systems (CLOUDS)
Laboratory at the University of Melbourne, Australia.  His research interests include distributed systems and cloud computing. He is currently working on resource management techniques using  data-driven  methods to optimise the  energy and thermal aspects of large-scale cloud data center resoruces.
\end{IEEEbiography}
\vspace{-0.5in}
\begin{IEEEbiography}
 [{\includegraphics[width=1in,height=1in,clip,keepaspectratio]{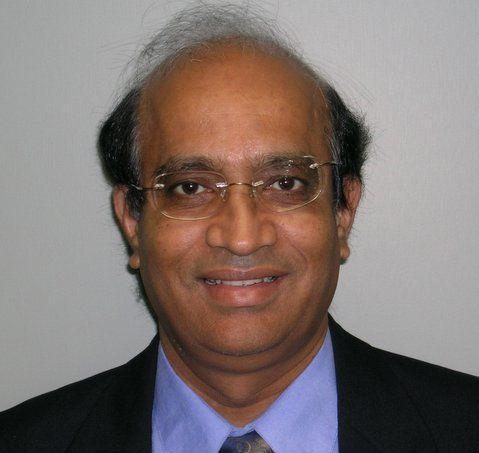}}]
    {Kotagiri Ramamohanarao }
received the PhD degree from Monash University. He is currently a professor of computer science with the University of Melbourne. He served on the editorial boards of the Computer Journal. At present, he is on the editorial boards of Universal Computer Science, Data Mining, and the International Very  Large Data Bases Journal. He was the program co-chair for VLDB, PAKDD, DASFAA, and DOOD conferences.
\end{IEEEbiography}
\vspace{-0.5in} 
\begin{IEEEbiography}
  [{\includegraphics[width=1in,height=1in,clip,keepaspectratio]{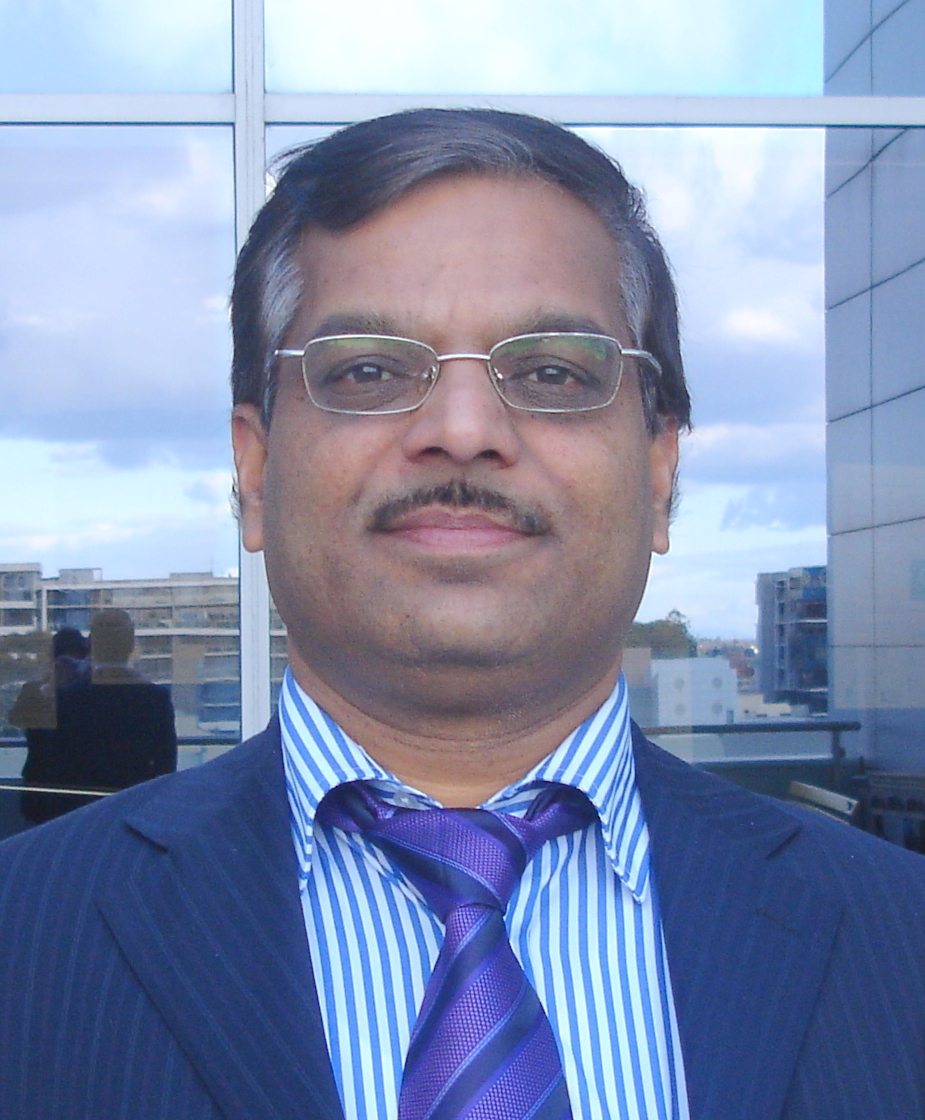}}]
    {Rajkumar Buyya} is a Redmond Barry Distinguished Professor and Director of the Cloud Computing and Distributed Systems (CLOUDS) Laboratory at the University of Melbourne, Australia. He is also serving as the founding CEO of Manjrasoft, a spin-off company of the University, commercializing its innovations in cloud computing. He has authored over 625 publications and seven textbooks including "Mastering Cloud Computing" published by McGraw Hill, China Machine Press, and Morgan Kaufmann for Indian, Chinese and international markets respectively. He is one of the highly cited authors in computer science and software engineering worldwide (h-index=138, g-index=307, 103400+ citations).  He is a fellow of the IEEE.
\end{IEEEbiography}

\end{document}